\documentclass[12pt]{amsart}
\usepackage{amsmath}
 \usepackage[foot]{amsaddr}

\usepackage{tikz}
\usetikzlibrary{matrix,arrows}
\usepackage{amsfonts,amssymb,amsthm, txfonts, pxfonts,amscd}
\usepackage[stretch=10]{microtype}
\def\struckint{\mathop{%
\def\mathpalette##1##2{\mathchoice{##1\displaystyle##2}%
 {##1\textstyle##2}{##1\scriptstyle##2}{##1\scriptscriptstyle##2}}%
\mathpalette
{\vbox\bgroup\baselineskip0pt\lineskiplimit-1000pt\lineskip-1000pt
\halign\bgroup\hfill$}
{##$\hfill\cr{\intop}\cr\diagup\cr\egroup\egroup}%
}\limits}
\usepackage[numbers]{natbib}
\usepackage{color}
\usepackage{booktabs,caption,fixltx2e}
\usepackage[flushleft]{threeparttable}
\usepackage{amsmath}
\usepackage{tikz}
\usetikzlibrary{matrix,arrows}
\usepackage{amsfonts,amssymb,amsthm, txfonts, pxfonts,amscd}
\usepackage{multirow}
\usepackage{floatrow}
\usepackage{subcaption}
\newfloatcommand{capbtabbox}{table}[][\FBwidth]

\usepackage[margin=1in]{geometry}
\usepackage{setspace}
\usepackage{soul,color, mathtools, fixmath}
\usepackage{multirow}
\usepackage[shortlabels]{enumitem}

\def\E{\mathcal{E}}

\newtheorem{thm}{Theorem}[section]
\newtheorem{lemma}[thm]{Lemma}

\newtheorem{corollary}[thm]{Corollary}

\newtheorem{assumption}[thm]{Assumption}
\newtheorem{remark}[thm]{Remark}%\endlocaldefs
\makeatletter

\def\indep{\mathrel{\rlap{$\perp$}\kern1.6pt\mathord{\perp}}}

\def\E{\mathbb{E}}
\def\P{\mathbb{P}}

\def\given{\, | \,}

\def\dotminussym#1#2{%
  \setbox0=\hbox{$\m@th#1-$}%
  \kern.5\wd0%
  \hbox to 0pt{\hss\hbox{$\m@th#1-$}\hss}%
  \raise.6\ht0\hbox to 0pt{\hss$\m@th#1.$\hss}%
  \kern.5\wd0}

\mathchardef\mhyphen="2D
\begin{document}

\title{Assessing Time-Varying Causal Effect Moderation in the Presence of Cluster-Level Treatment Effect Heterogeneity}
\author{Jieru Shi}
\author{Zhenke Wu}
\author{Walter Dempsey}
\address {Department of Biostatistics, University of Michigan, 1415 Washington Heights, Ann Arbor, MI 48109, USA}
\email{wdem@umich.edu}

\date{\today}

\begin{abstract}
The micro-randomized trial (MRT) is a sequential randomized experimental design to empirically evaluate the effectiveness of mobile health (mHealth) intervention components that may be delivered at hundreds or thousands of decision points. MRTs have motivated a new class of causal estimands, termed ``causal excursion effects", for which semiparametric inference can be conducted via  a weighted, centered least squares criterion (Boruvka et al., 2018). Existing methods assume between-subject independence and non-interference. Deviations from these assumptions often occur.
% which, if unaccounted for, may result in bias and overconfident variance estimates.
In this paper, causal excursion effects are revisited under potential cluster-level treatment effect heterogeneity and interference, where the treatment effect of interest may depend on cluster-level moderators. Utility of the proposed methods is shown by analyzing data from a multi-institution cohort of first year medical residents in the United States.
\end{abstract}


\maketitle

\section{Introduction}

Modern behavioral science has placed a considerable amount of attention on push notifications sent via mobile device that are adapted to continuously collected information on an individual's current context.  These time-varying adaptive interventions are hypothesized to lead to meaningful short- and long-term behavior change. The assessment of the time-varying effect of such push notifications motivated sequential randomized designs such as the micro-randomized trial (MRT) \citep{Nahum2017, KlasnjaMRT}, in which individuals are randomized to potentially receive notifications at hundreds or thousands of decision points. The MRT design enables the estimation of marginal treatment effects of push notifications on pre-specified time-lagged outcomes of interest, referred to as ``causal excursion effects" \citep{Boruvkaetal, Qian2021, DempseyAOAS}. Semiparametric inference of the causal excursion effects can be conducted via a weighted, centered least squares (WCLS) criterion \citep{Boruvkaetal}.
%The WCLS approach has also been used in analyzing data obtained from stratified MRT~\citep{DempseyAOAS}.

The WCLS inferential method relies on two key assumptions. First, an intervention delivered to an individual is assumed to only impact that same individual's  outcomes, i.e., between-subject non-interference. Second, the method assumes no stochastic dependence among outcomes of different subjects. Deviations from these assumptions, however, may occur when individuals naturally form clusters. To address these violations, this paper extends the definition of causal excursion effects to account for potential interference and cluster-level treatment effect heterogeneity accompanied with a general inferential approach that provides valid inferences and subsumes WCLS as a special case.
% We propose and illustrate a novel inferential approach that ensures proper interval coverage.

\section{Preliminaries}
\label{section:preliminaries}

\subsection{Micro-Randomized Trials (MRT)}

An MRT consists of a sequence of within-subject decision times $t=1,\ldots,T$ at which treatment options may be randomly assigned~\citep{Liaoetal2015}.  Individual-level data can be summarized as~$\{ O_0, O_1, A_1, O_2, A_2, \ldots, O_T, A_T, O_{T+1} \}$
where $t$ indexes a sequence of decision points, $O_0$ is the baseline information, $O_t$ is the information collected between time $t-1$ and $t$, and $A_t$ is the treatment option provided at time $t$; for simplicity, we consider binary treatment options, i.e.,~$A_t \in \{ 0, 1\}$.  In an MRT, $A_t$ is randomized according to a known sequence of randomization probabilities that may depend on the complete observed history $H_t := \{ O_0, O_1, A_1, \ldots, A_{t-1}, O_t \}$,  denoted ${\bf p} = \{ p_u (A_u \given H_u) \}_{u=1}^t$. Treatment options are designed to impact a proximal response, denoted by $Y_{t,\Delta}$, which is a known function of the participant’s data within a subsequent window of length~$\Delta \ge 1$, i.e., $Y_{t,\Delta}=y(O_t, A_t, O_{t+1},A_{t+1},\ldots, A_{t+\Delta-1},O_{t+\Delta})$~\citep{DempseyAOAS}.
% ; when $\Delta=1$, we set $Y_{t,1}=y(O_t, A_t, O_{t+1})$.

\subsection{Estimand and Inferential Method: A Review}
\label{section:standardmrtmethods}

% NEED CLEANER INTRO INTO DELTA ISSUE
% COULD DO POTENTIAL OUTCOME QUICKLY?
% In this paper, we focus on presenting the causal analysis with time lag $\Delta=1$; the framework is general and applies to an arbitrary number of lags $\Delta \geq 1$ \citep{Boruvkaetal}. This section assumes stochastic independence between subjects and briefly reviews the existing estimand and inferential precedure.
We focus on the class of estimands referred to as ``causal excursion effects", which are time-varying as a function of the decision point~$t$. We provide formal definitions using potential outcomes~\citep{Rubin, Robins}.  Let~$Y_{t,\Delta} (\bar a_{t+\Delta-1})$ denote the potential outcome for the proximal response under treatment sequence~$\bar a_{t+\Delta-1} = (a_1, \ldots, a_{t+\Delta-1})$.  Let $S_t (\bar a_{t-1})$ denote the potential outcome for a potential time-varying effect moderator which is a deterministic function of the potential history up to time $t$, $H_t (\bar a_{t-1})$. The causal excursion effect is then defined with respect to a \emph{reference distribution}, i.e., the distribution of treatments $\bar A_{t+\Delta-1} := \{A_1, \ldots, A_{t+\Delta-1}\}$. %from time $1$ to ~$t+\Delta-1$.
For past treatments, $\bar A_t$, we follow common practice in observational mobile health studies where analyses such as GEEs~\citep{Liang1986} are conducted marginally over the distribution of historical information.  A similar strategy here is to use the past treatment randomization probabilities as the reference distribution between time $1$ and $t$.  For future treatments, the choice of distribution for $\tilde A_{t+1:(t+\Delta-1)} := (A_{t+1}, \ldots, A_{t+\Delta-1})$ may differ by the type of inference desired; note that when $\Delta=1$, future treatments do not impact the proximal outcome and therefore a reference distribution for future treatments is not necessary.
%because proximal response of interest $Y_{t,1}$ is prior to any future randomizations
Here, we assume the reference distribution for treatment assignments from $t+1$ to $t+\Delta-1$ is given by a randomization probability generically represented by~$\pi_{u}(a_{u} | H_{u}), u=t+1,\ldots, t+\Delta-1$ and let $\pi=\{\pi_{u}\}_{u=t+1}^{t+\Delta-1}$.  This generalization contains previous definitions such as lagged effects~\citep{Boruvkaetal} where $\pi_{u} = p_{u}$ and deterministic choices such as $a_{t+1:(t+\Delta-1)} = {\bf 0}$~\citep{DempseyAOAS, Qian2021} where $\pi_{u} = {\bf 1}\{a_{u} = 0\}$ and $1\{\bullet\}$ is the indicator function.  Then the causal excursion effect $\beta_{{\bf p}, \pi, \Delta} (t;s)$ is defined as

\begin{align}
&\E_{{\bf p}, \pi} \left [ Y_{t,\Delta} \left(\bar A_{t-1}, 1, \tilde A_{t+1:(t+\Delta-1)} \right) - Y_{t,\Delta} \left(\bar A_{t-1}, 0, \tilde A_{t+1:(t+\Delta-1)} \right) \given S_t (\bar A_{t-1}) = s\right] \label{eq:causalexcursion_po} \\
=  &\E_{{\bf p}} \left[ \E_{{\bf p}} \left[ W_{t,\Delta} Y_{t,\Delta} \mid A_t = 1, H_t \right] - \E_{{\bf p}} \left[ W_{t,\Delta} Y_{t,\Delta} \mid A_t = 0, H_t \right] \mid S_t = s \right] \label{eq:causalexursion}
\end{align}
where treatment sequence up to time $t-1$: $\bar A_{t-1} \sim {\bf p}$, future treatment sequence up to $t+\Delta-1$: $\tilde A_{t+1:(t+\Delta-1)} \sim \pi$ and~$W_{t,\Delta} = \prod_{u=t+1}^{t+\Delta-1} \pi_u (A_u | H_u) / p_u(A_u | H_u)$ can be interpreted as change of measure from ${\bf p}$ to $\pi$ for treatment assignments $\tilde A_{t+1:(t+\Delta-1)}$; we set $W_{t,\Delta}=1$ when $\Delta=1$.
Equation~\eqref{eq:causalexursion} expresses \eqref{eq:causalexcursion_po} in terms of observable data, which requires the standard causal inference assumptions of positivity, sequential ignorability, and consistency.
Assuming $\beta_{{\bf p}, \pi, \Delta} (t;s) = f_t(s)^\top \beta^\star$ where $f_t(s) \in \mathbb{R}^q$ is a feature vector comprised of a $q$-dimensional summary of observed state information depending only on state $s$ and decision point $t$, a consistent estimator for $\beta^*$ can be obtained by minimizing a weighted and centered least squares (WCLS) criterion:
\begin{equation}
\label{eq:mrtstandard}
\hat \beta = \arg \min_{\alpha, \beta} \mathbb{P}_n \left[ \sum_{t=1}^T W_t \times W_{t,\Delta} \left[ Y_{t,\Delta} - g_t(H_t)^\top \alpha - \left ( A_t - \tilde p_t (1 \mid S_t) \right) f_t (S_t)^\top \beta \right]^2 \right]
\end{equation}
where~$\mathbb{P}_n$ is shorthand for the sample average, $W_t = \tilde p_t (A_t \mid S_t) / p_t (A_t \mid H_t)$ is a weight where the numerator is an arbitrary function with range $(0,1)$ that only depends on potential moderators of interest $S_t$, and $g_t(H_t) \in \mathbb{R}^p$ are $p$ control variables chosen to help reduce variance and to construct more powerful test statistics. See \citep{Boruvkaetal} for more details on the seminal estimand formulation and consistency, asymptotic normality, and robustness properties of the WCLS estimation method.
% In HeartSteps, for example, a natural control variable is the number of steps in the prior 30 minutes which is likely highly correlated with the proximal response and thus can be used to reduce variance in the estimation of $\beta$. See~\cite{Boruvkaetal} for estimation, consistency, and small-sample corrected uncertainty assessment that is robust to the independent working correlation assumption in~\eqref{eq:mrtstandard}.

\subsection{Motivating Example}
\label{section:motex}

\begin{figure}
  % \figuresize{.3}
  % \figurebox{10pc}{15pc}{}[heterogeneityv2.eps]
  \includegraphics[width = 0.7\textwidth]{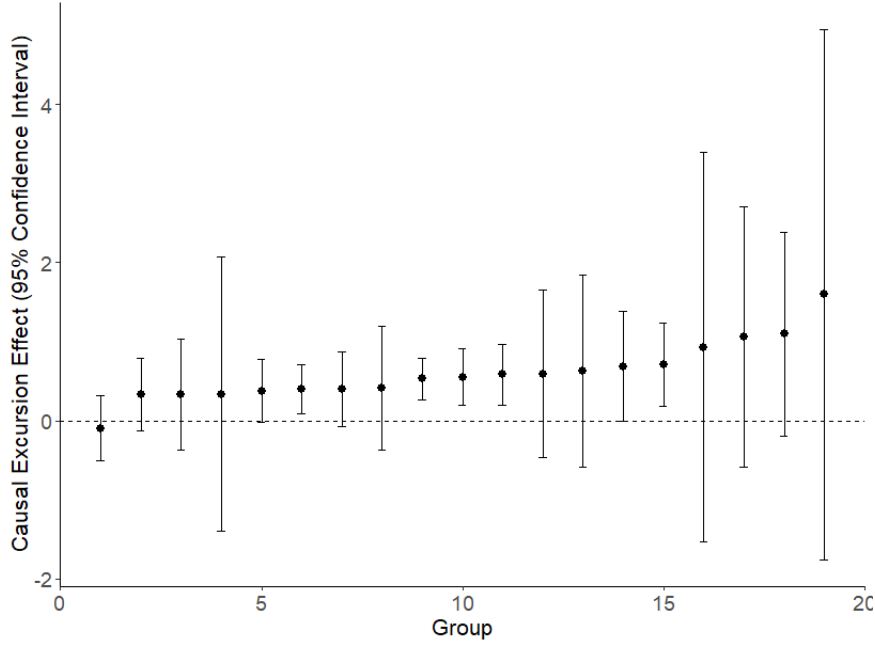}
  \caption{Heterogeneous estimated causal excursion effects across the largest $M=19$ clusters defined as specialties of size $G \geq 6$; the estimates are obtained using cluster-specific applications of the existing WCLS method.}
  \label{fig:wcls_heterogeneity}
\end{figure}

The Intern Health Study (IHS) is a 6-month MRT on 1,562 medical interns~\citep{Necamp2020}.  Due to high depression rates and levels of stress during the first year of physician residency training, a critical question is whether targeted notifications can improve mood, increase sleep time, and/or increase physical activity. Enrolled medical interns were randomized weekly to receive either mood, activity, or sleep notifications or receive no notifications for that week (probability 1/4 each).  Analyses conducted in this paper focus on the weekly randomization; see~\cite{Necamp2020} for further study details.  Figure~\ref{fig:wcls_heterogeneity} presents specialty-specific effect estimates on weekly average mood scores using~\eqref{eq:mrtstandard}, with evident specialty-level treatment effect heterogeneity.  This suggests a marginal analysis must account for effect heterogeneity at the specialty-level. The present work offers such a framework to address the inferential deficiency that we show if the standard WCLS were used under cluster-level effect heterogeneity. In addition, there exists potential within-cluster interference of other subjects' treatments upon a subject's outcome. Our framework defines a new and useful indirect excursion effect under sequential treatments in contrast to existing work that mostly focuses on indirect effects in non-temporal settings.

\section{Cluster-Level Proximal Treatment Effects}
\label{section:cond_effects}

\subsection{Proximal Moderated Treatment Effects: A Cluster-based Conceptualization}
\label{section:prox_effects_pot_outcome}

Consider a cluster of size $G$.  Overbar will continue to denote treatment sequences; $\bar a_{t,j} = (a_{1,j},\ldots, a_{t,j})$, for instance, denotes the sequence of realized treatment sequence up to and including decision time $t$ for individual $j \in [G]:=\{1,\ldots, G\}$.   Let $\bar a_{t} = (\bar a_{t,1}, \ldots, \bar a_{t,G})$ denote the set of realized treatments for all individuals in the cluster. Let $\bar a_{t,-j} = \bar a_t \backslash \bar a_{t,j}$ denote this set with the $j$th individual removed. Let $Y_{t,\Delta,j} (\bar a_{t+\Delta-1})$ denote the potential outcome for individual $j \in [G]$ which may depend on realized treatments for all subjects in the cluster.
% and that cluster-size may vary.
% \zw{cluster sizes may differ}

\noindent \textbf{Direct causal excursion effects}.
In standard MRTs, the individual is the unit of interest.  Here, the cluster is the unit of interest and the effect of interest is in providing treatment versus not providing treatment at time $t$ on a random individual in the group.  This can be expressed as a difference in potential outcomes for the proximal response
\begin{equation}
\label{eq:directeffect}
\begin{array}{r@{}l}
\frac{1}{G} \sum_{j=1}^G \bigg [ Y_{t, \Delta,j} &(\bar a_{t+\Delta-1,-j}, (\bar a_{t-1,j}, 1, a_{t+1:(t+\Delta-1),j})) \\
&- Y_{t,\Delta, j} (\bar a_{t+\Delta-1,-j}, (\bar a_{t-1,j}, 0, a_{t+1:(t+\Delta-1),j})) \bigg].
\end{array}
\end{equation}
Following~\cite{Halloran1995} and~\cite{Tchetgen2012}, \eqref{eq:directeffect} is a \emph{group average direct causal effect} of treatment versus no treatment fixing all other treatments.
% \zw{clarify the directness means blocking all other paths except the path from $A_{tj}$ to $Y_{t+1,j}$}

The ``fundamental problem of causal inference''~\citep{Rubin, Pearl2009} is that individual differences cannot be observed. Thus, similar to prior work~\citep{DempseyAOAS, Boruvkaetal}, averages of potential outcomes are considered. Let $S_t (\bar a_{t-1})$ denote a vector of potential moderator variables formed from $H_t (\bar a_{t-1})$, the \emph{cluster-level} history up to decision point $t$.
% We write $S_t (\bar a_{t-1}) = \left( S_{t,j} (\bar a_{t-1}), S_{t,-j} (\bar a_{t-1}) \right)$ to clarify that the potential moderator variables can contain both information on the selected individual as well as other individuals in the cluster.
Then the moderated direct treatment effect, denoted $\beta_{{\bf p}, \pi, \Delta} (t; s)$, can be defined as
\begin{equation}
\label{eq:directavglineareffect}
\begin{array}{r@{}l}
\mathbb{E}_{{\bf p}, \pi} \bigg[ &Y_{t,\Delta,J} (\bar A_{t+\Delta-1,-J}, (\bar A_{t-1,J}, 1, \tilde A_{t+1:(t+\Delta-1),J} )) \\
- &Y_{t,\Delta,J} (\bar A_{t+\Delta-1,-J}, (\bar A_{t-1,J}, 0,\tilde A_{t+1:(t+\Delta-1), J})) \mid  S_{t} (\bar A_{t-1}) = s \bigg].
\end{array}
\end{equation}
where $J$ is a uniformly distributed random index defined on $[G]$. The expectation is over the potential outcomes~$Y_{t,\Delta,J}(\cdot)$, the randomized treatments -- $\bar A_{t+\Delta-1,-J} \sim {\bf p}$, $\bar A_{t-1,J} \sim {\bf p}$, and $\tilde A_{t+1:(t+\Delta-1),J} \sim \pi$ -- and the random index~$J$.
% \zw{Perhaps can talk about the expectation is over both Y and A; and write it in a way that addresses potential issues that may be raised by the utility of this estimand and how is it different from existing ones.}  \wd{Now raising the expectations point but pushing discussion of the estimand and its difference from other stuff to that 3.2.3 section}
Choice of $S_t (\bar A_{t-1})$ depends on the scientific question of interest.  A primary analysis may focus on marginal effects and set $S_t (\bar A_{t-1}) = \emptyset$.  A second analysis may focus on assessing the effect conditional on variables only related to the individual indexed by $J$ and set $S_t (\bar A_{t-1}) = X_{t,J} (\bar A_{t-1,J})$, i.e., a potential individual-level moderator of interest where $X_{t,J}(\bar{A}_{t-1,J})$ produces a vector of summary variables from the history of individual $J$, $H_{t,J}(\bar{A}_{t-1,J})$.  A third analysis may consider group-level moderators such as $S_t (\bar A_{t-1}) = G^{-1} \sum_j X_{t,j} (\bar A_{t-1,j})$ or $S_t (\bar A_{t-1}) = \left( X_{t,j}(\bar A_{t-1,j}), \frac{1}{G-1} \sum_{j^\prime \neq j} X_{t,j^\prime} (\bar A_{t-1,j^\prime}) \right)$. Equation~\eqref{eq:directavglineareffect} generalizes the \emph{population average direct causal effect} from~\cite{Tchetgen2012} to a \emph{group-level causal excursion effect} that allows for moderation and time-varying treatments.

\noindent \textbf{Pairwise indirect causal excursion effects}.
Of secondary interest is the indirect effect of providing treatment versus not providing treatment to the $j$th individual at time $t$ on a different individual's proximal response, i.e., pairwise within-cluster treatment interference.  Here, we define the  \emph{pairwise indirect causal excursion effect} as
\begin{align}
\frac{1}{G \cdot (G-1)} &\sum_{j^\prime \neq j} \left[Y_{t,\Delta,j} (\bar a_{t+\Delta-1,-\{j, j^\prime\}}, (\bar a_{t-1,j}, 0, a_{t+1:(t+\Delta-1),j}), (\bar a_{t-1,j^\prime}, 1, a_{t+1:(t+\Delta-1),j'})) \right.\nonumber\\
&- \left.Y_{t,\Delta,j} (\bar a_{t+\Delta-1,-\{j, j^\prime\}}, (\bar a_{t-1,j}, 0, a_{t+1:(t+\Delta-1),j}), (\bar a_{t-1,j^\prime}, 0, a_{t+1:(t+\Delta-1),j'})) \nonumber \right].
\end{align}
Again, since individual differences cannot be observed, averages of potential outcomes are considered. The moderated pairwise indirect treatment effect, denoted $\beta^{(IE)}_{{\bf p}, \pi, \Delta} (t; s)$, is
\begin{equation}
\label{eq:inddirectavglineareffect}
\begin{array}{r@{}l}
&\mathbb{E}_{{\bf p}, \pi} \bigg[ Y_{t,\Delta,J} \left(\bar A_{t+\Delta-1,-\{J, J^\prime\}}, (\bar A_{t-1,J}, 0, \tilde A_{t+1:(t+\Delta-1),J}), (\bar A_{t-1,J'}, 1, \tilde A_{t+1:(t+\Delta-1),J'}) \right) \\
&- Y_{t,\Delta,J} \left(\bar A_{t+\Delta-1,-\{J, J^\prime\}}, (\bar A_{t-1,J}, 0, \tilde A_{t+1:(t+\Delta-1),J}), (\bar A_{t-1,J'}, 0, \tilde A_{t+1:(t+\Delta-1),J'}) \right) \mid S_t (\bar A_{t-1}) =s\bigg].
\end{array}
\end{equation}
where $J^\prime$ is uniformly distributed random index on the set $[G] \backslash \{J\}$. The expectation is over both the potential outcomes~$Y_{t,\Delta,J}(\cdot)$, randomized treatments -- $\bar A_{t+\Delta-1,-\{J,J^\prime\}} \sim {\bf p}$, $\bar A_{t-1,J} \sim {\bf p}$, $\bar A_{t-1,J^\prime} \sim {\bf p}$, $\tilde A_{t+1:(t+\Delta-1),J} \sim \pi$, and $\tilde A_{t+1:(t+\Delta-1),J^\prime} \sim \pi$ --  and the random indices~($J$ and $J^\prime$). The potential moderator can be written as $S_t (\bar A_{t-1}) = \left( S_{t,J} (\bar A_{t-1}), S_{t,J^\prime} (\bar A_{t-1}), S_{t,-\{J, J^\prime \}} (\bar A_{t-1}) \right)$ to clarify that the variables can contain both information on the two selected individuals as well as others in the cluster. Another pairwise indirect effect can be defined when individual~$J$ receives treatment, i.e., $A_{t,J} = 1$ instead of~$A_{t,J} = 0$ as in~\eqref{eq:inddirectavglineareffect}.

\begin{remark}
The effect defined by~\eqref{eq:inddirectavglineareffect} generalizes the \emph{group average indirect causal effect} from~\cite{Tchetgen2012} to a \emph{group-level pairwise indirect causal excursion effect} that allows for moderation and time-varying treatments. To see this, note that the excursion effect at each decision time~$t$ averages over a particular reference distribution over the past and future treatments up to and including time~$t+\Delta-1$ defined by the MRT randomization probabilities~$\bf p$ and the alternative probability distribution~$\pi$.  The contrast is over two treatment allocations, both where a random individual does not receive treatment, but where in one allocation another random individual receives treatment and in the other allocation that same individual does not receive treatment. \cite{Tchetgen2012} consider contrasts between any two randomized treatment allocations conditional on a random individual not receiving treatment in a non-temporal setting.  Therefore, per decision time our definition is a special case of their indirect effect where, for a random non-treated individual, we focus on treating or not treating another random individual and marginalizing over all others in the group.  More complex contrasts could be derived such as three- or four-way indirect effects; however, the number of combinations grows quickly making estimation unrealistic in our setting. Our choice of contrast was thus motivated by finding an estimand of scientific interest which could be reasonably estimated within the MRT setting, bridging the literature on causal excursions and indirect effects.
% so~\eqref{eq:inddirectavglineareffect} can be seen as a generalization to the time-varying treatment setting.
\end{remark}

\subsection{Causal Excursion Effect Estimand Depends on Treatment Distribution}

% It is useful to distinguish the above estimands, the traditional MRT estimand~\citep{Boruvkaetal}, and estimands commonly studies in longitudinal treatment effect estimation~\citep{Robins}. In the causal inference literature, the typical estimand is an expected outcome for a particular sequence of treatments,i.e.,~$\E \left[ Y(a_1,\ldots, a_T)\right]$.  Such estimands do not depend on the treatment distribution, but are often not of primary interest in the current MRT setting since many sequences of treatment may never be observed in finite samples given the large number of decision points (often hundreds or thousands).

Estimands considered here are most similar to average outcomes under a particular dynamic treatment regime~$\E_\mu \left[ Y(A_1,\ldots, A_T)\right]$ where $\mu$ denotes the dynamic treatment regime from which the treatments are drawn~\citep{MurphyJASA2001}. Indeed, for any $A_{u,j}$ not contained in $S_t (\bar A_{t-1})$, the direct and indirect effects depend on the distribution of $\{ A_{u,j} \}_{u \le t + \Delta -1, j \in [G]}$.  Estimands~\eqref{eq:directavglineareffect} and~\eqref{eq:inddirectavglineareffect} marginalize over treatments not contained in $S_t(\bar A_{t-1})$.  Marginalization over different probabilistic assignment of treatments may yield different results. Therefore, the direct and indirect excursion effects depend on the study protocol and choice of alternative distribution~$\pi$. The reason for this is that micro-randomization is meant to gather information on how to optimize the design of intervention components~\citep{Collins2018}.  The marginal formulation of main and moderation effects contrasts excursions from the current treatment protocol, and mimics analyses used in a factorial design that marginalize over factors including time. See~\citep[Section 8]{Qian2021} for additional considerations.  Regardless, the effects considered in this paper are causal and depend on the treatment assignment distributions.  Due to this dependence, in real data analysis, we recommend presenting the micro-randomization distribution together with the estimated treatment effects, thus the subscript~$({\bf p}, \pi)$ in the definition of both the direct and indirect effects are specified.

\subsection{Identification}
\label{section:prox_effects_data}
Causal effects \eqref{eq:directavglineareffect} and~\eqref{eq:inddirectavglineareffect} can be expressed in terms of the observable data under the following standard set of causal inference assumptions~\citep{Robins}:

\begin{assumption} \normalfont
  \label{consistency}
  We assume consistency, positivity, and sequential ignorability:
  \begin{itemize}
  \item Consistency: For each~$t \leq T$ and $j \in [G]$,
    $\{Y_{t,\Delta,j} (\bar{A}_{t+\Delta-1} ), O_{t,j} (\bar A_{t-1}), A_{t,j} (\bar{A}_{t-1} )\}  = \{Y_{t, \Delta, j}, O_{t,j}, A_{t,j}\}$, i.e., observed values equal the corresponding potential outcomes;
  \item Positivity: if the joint density~$\{ A_t = a_t, H_t = h_t\}$ is greater
    than zero, then~$P (A_t = a_t \given H_t = h_t ) > 0$;
  \item Sequential ignorability: for each~$t \leq T$, the
    potential outcomes,\\ $\{Y_{2,\Delta,j} ( \bar a_{1+\Delta-1}), O_{2,j}(a_{1}),A_{2,j}( a_{1}), \ldots,
    Y_{T,\Delta, j} (\bar a_{T+\Delta-1}) \}_{j \in [G], \bar a_{T+\Delta-1}\in \{0,1\}^{(T+\Delta-1)\times G}}$, are independent of~$A_{t,j}$ conditional on the observed history~$H_t$.
  \end{itemize}
\end{assumption}

Sequential ignorability and, assuming all of the randomization probabilities are bounded away from $0$ and $1$, positivity, are guaranteed in our setting  by design.
% In a standard MRT, the randomization probabilities may depend on the individual's observed history so $\Pr(A_t = a_t \mid H_t = h_t) = \prod_{j=1}^G \Pr(A_{t,j} = a_{t,j} \mid H_{t,j} = h_{t,j})$ and the positivity constraint can be placed on the individual-level randomization probabilities.  Here, we allow for the possibility that the randomization probabilities depend on the cluster-level history.
Consistency is a necessary assumption for linking the potential outcomes as defined here to the data. Since an individual's outcomes may be influenced by the treatments provided to other individuals in the same cluster, consistency holds due to our use of a cluster-based conceptualization of potential outcomes as seen in~\cite{Hong2006} and \cite{Vanderweele2013}.

\begin{lemma}
\label{lemma:cond_effect}
  Under Assumption~\ref{consistency}, the moderated direct treatment effect~$\beta_{{\bf p}, \pi, \Delta} (t;s)$ is equal to
  $$
  \mathbb{E} \left[ \mathbb{E} \left[ W_{t, \Delta, J} Y_{t,\Delta,J} \mid H_{t}, A_{t,J} = 1 \right] - \mathbb{E} \left[ W_{t, \Delta, J} Y_{t, \Delta,J} \mid H_{t}, A_{t,J} = 0 \right] \mid S_t=s \right],
  $$
  where expectations are with respect to the distribution of the
  data collected under the actual treatment assignment probabilities ${\bf p}$, and~$W_{t,\Delta,j} = \prod_{u=t+1}^{t+\Delta-1} \pi(A_{u,j} | H_u) / p(A_{u,j} | H_u)$ with $W_{t,1,j}=1$; and the moderated indirect treatment effect~$\beta^{(IE)}_{{\bf p}, \pi, \Delta}(t;s) $ is equal to
  \begin{align*}
  \mathbb{E} \bigg[ \mathbb{E} &\left[ W_{t,\Delta,J,J'} Y_{t,\Delta,J} \mid H_{t}, A_{t,J} = 0, A_{t,J^\prime} = 1 \right] \\
  - &\mathbb{E} \left[ W_{t,\Delta,J,J'} Y_{t,\Delta,J} \mid H_{t}, A_{t,J} = 0, A_{t,J^\prime} = 0 \right] \mid S_t=s \bigg].
  \end{align*}
\noindent where~$W_{t,\Delta,j,j'} = \prod_{u=t+1}^{t+\Delta-1} \pi_u(A_{u,j}, A_{u,j'} | H_u) / p_u(A_{u,j}, A_{u,j'} | H_u)$ with $W_{t,1,j,j'}=1$.
\end{lemma}
\noindent Proof of Lemma~\ref{lemma:cond_effect} can be found in the Appendix~\ref{app:techdetails}.

\section{Estimation}
\label{section:estimation}

% Motivated by the identification result in Lemma~\ref{lemma:cond_effect}, we consider estimation of direct and indirect effects along with uncertainty assessments.

\subsection{Direct Causal Excursion Effect Estimation}

\begin{assumption} \normalfont
\label{ass:directeffect}
Assume the direct causal excursion effect~$\beta_{{\bf p}, \pi, \Delta}(t;s) = f_t (s)^\top \beta^\star$ where $f_t (s) \in \mathbb{R}^q$ is a $q$-dimensional feature vector that is a function of moderator state $s$ and decision point $t$. \end{assumption}

Consider inference on the $q$-dimensional parameter $\beta^\star$. Define the weight~$W_{t,j}$ at decision time $t$ for the $j$th individual as equal to $\frac{\tilde p_t (A_{t,j} \mid S_t)}{p_t (A_{t,j} \mid H_t)}$ where $\tilde p_t (a \mid S_t)\in (0,1)$ is arbitrary as long as it does not depend on terms in $H_t$ other than $S_t$, and $p(A_{t,j} \mid H_t)$ is the marginal probability that individual $j$ receives treatment $A_{t,j}$ given $H_t$.
% For now, we consider pre-specified $\tilde p_t (a \mid S_t)$  (i.e., does not depend on the observed MRT data).
Here we consider an estimator which is the minimizer of a cluster-based, weighted-centered least-squares (C-WCLS) criterion:
\begin{equation}
\label{eq:directwcls}
\mathbb{P}_M \left[ \frac{1}{G_m} \sum_{j=1}^{G_m} \sum_{t=1}^T W_{t,j} \times W_{t,\Delta,j} \left(Y_{t,\Delta,j} - g_t (H_t)^\top \alpha - (A_{t,j} - \tilde p_t (A_{t,j} \mid S_t)) f_t(S_t)^\top \beta \right)^2 \right]
\end{equation}
where $\mathbb{P}_M$ is defined as the average of a function over the sample, which in this context is the sample of \emph{clusters} rather than the sample of \emph{individuals} as in traditional MRT settings. In Appendix~\ref{app:asymptotics}, we prove the following result.

\begin{lemma}
\label{lemma:asymnorm}
Under Assumption~\ref{ass:directeffect}, given invertibility and moment conditions, the estimator $\hat \beta$ that minimizes \eqref{eq:directwcls} satisfies $\sqrt{M} \left(\hat \beta - \beta^\star \right) \to N(0, Q^{-1} W Q^{-1})$ where
$$
Q = \mathbb{E} \left[ \sum_{t=1}^T \tilde p_t( 1 \mid S_{t} ) ( 1- \tilde p_t( 1 \mid S_{t} )) f_t (S_t) f_t (S_t)^\top \right]
$$
and
\begin{align*}
W =  \mathbb{E} \bigg[ \sum_{t=1}^T &W_{t,J} \times W_{t,\Delta,J} \, \epsilon_{t,J} ( A_{t,J} - \tilde p_t( 1 \mid S_{t} )) f_t (S_t) \\
\times \sum_{t=1}^T &W_{t, \tilde J} \times W_{t,\Delta,\tilde J} \, \epsilon_{t, \tilde J} ( A_{t, \tilde J} - \tilde p_t( 1 \mid S_{t} )) f_t (S_t)^\top  \bigg],
\end{align*}
where $\epsilon_{t,j} = Y_{t,\Delta, j} - g_t(H_t)^\top \alpha^\star - (A_{t,j} - \tilde p_t (1 \mid S_t) ) f_t (S_t)^\top \beta^\star$, $\alpha^\star$ minimizes the least-squares criterion $\mathbb{E}  \left[G_m^{-1} \sum_{j=1}^{G_m} \sum_{t=1}^T W_{t,j} W_{t,\Delta,j} \left( Y_{t,\Delta,j} - g_t(H_t)^\top \alpha \right)^2 \right]$, and both $J$ and $\tilde J$ are independent randomly sampled indices from the same cluster.
\end{lemma}

\noindent In practice, plug-in estimates $\hat Q$ and $\hat W$ are used to estimate the covariance structure; Appendix~\ref{app:ssa} presents their estimates with small-sample adjustments.

% It is clear from Lemma~\ref{lemma:asymnorm} that standard MRT data analytic methods, i.e., equation~\eqref{eq:mrtstandard}, will produce biased estimates for the marginal effect of interest if cluster size is variable.

\begin{remark}{\it ($L_2$ Projection Interpretation)}
\label{rmk:l2proj}
Importantly, Assumption~\ref{ass:directeffect} is not required.  That is, we can follow~\cite{Neugebauer2007,Rosenblum2010,Kennedy2019,DempseyAOAS} and others in using $f_t (s)^\top \beta$ as a working model for $\beta_{{\bf p}, \pi, \Delta} (t;s)$.  Specifically,~$\hat \beta$ is a solution to the weighted least-squares projection
$$
\beta^\star = \arg \min_{\beta} \mathbb{E} \left[ \frac{1}{G} \sum_{j=1}^G \sum_{t=1}^T \tilde p_t(1 \mid S_t) \left( 1 - \tilde p_t(1 \mid S_t) \right) \left( \beta(t; S_t) - f_t (S_t)^\top \beta \right)^2 \right].
$$
Here, the weight is the variance of the numerator in the weight~$W_{t,j}$.  To interpret as a projection or as a correctly specified causal effect can be viewed as a bias-variance trade-off. The projection interpretation guarantees well-defined parameter interpretation in practice where Assumption~\ref{ass:directeffect} is unlikely to hold.  See~\citep[Sec. 3.1, pp.9--10][]{Kennedy2019} for a discussion of the use of projections in causal versus predictive settings.
\end{remark}

Appendix \ref{sec:semipar} presents semiparametric efficiency theory in the special case of $S_t = H_t$.

% Appendix \ref{section:samesies} presents conditions that guarantee the equivalence of C-WCLS to the standard WCLS approach in terms of effect estimates and asymptotic variance.

\subsection{Connection to the Standard MRT Analysis}
\label{section:samesies}
A natural question is whether there are conditions such that the standard MRT analysis presented in Section~\ref{section:standardmrtmethods} is equivalent to the proposed direct effect analysis.
Lemma~\ref{lemma:samesies} proves that, under certain conditions, an equivalence of estimates and asymptotic variances is guaranteed.

\begin{lemma}
\label{lemma:samesies}
Consider the direct effect when the moderator is defined on the individual (i.e., $S_{t,j}$), and the randomization probabilities only depend on the individual's observed history, i.e., $p(A_{t,j} \mid H_t) = p(A_{t,j} \mid H_{t,j})$.  If cluster size is constant (i.e., $G_m \equiv G$), then the point estimates from \eqref{eq:mrtstandard} and~\eqref{eq:directavglineareffect} are equal for any sample size. Moreover, if
\begin{equation}
    \label{eq:samesiescondition}
    \E \left[ \E \left[ W_{t,\Delta,j} \epsilon_{t,j} \times W_{t^\prime, \Delta, j^\prime} \epsilon_{t^\prime, j^\prime} \given H_{t,j}, A_{t,j}=a, H_{t^\prime,j^\prime}, A_{t^\prime,j^\prime} = a^\prime \right] \mid S_{t,j}, S_{t^\prime,j^\prime} \right]
\end{equation}
equals $\psi(S_{t,j}, S_{t^\prime,j^\prime})$
for some function $\psi$, i.e., the cross-terms are constant in $a$ and $a^\prime$, where $\epsilon_{t,j}$ is the error defined in Lemma~\ref{lemma:asymnorm}, then the estimators share the same asymptotic variance.
\end{lemma}

Proof of Lemma~\ref{lemma:samesies} can be found in Appendix~\ref{app:samesies}. Here, a class of random effect models is introduced to help with interpretation of the sufficient condition~\eqref{eq:samesiescondition}.  Specifically, for participant $j$ at decision time~$t$, let $\Delta =1$ and suppose the generative model for the proximal response is
$$
Y_{t,1,j} = g_t(H_{t,j})^\top \alpha + \underbrace{Z_{t,j}^\top b_g}_{(I)} + (A_{t,j} - p_t(1 \mid H_{t,j})) (f_t( H_{t,j})^\top \beta + \underbrace{Z_{t,j}^\top \tilde b_g}_{(II)}) + e_{t,j}
$$
where $(I)$ and $(II)$ are random effects with \textbf{}design matrix $Z_{t,j}$, $\E[ f_t(H_{t,j})^\top \beta \mid S_{t,j} ] = f_t(S_{t,j})^\top \beta$, and $e_{t,j}$ is a participant-specific error term.  The treatment effect conditional on the complete observed history and the random effects is $f_t( H_{t,j})^\top \beta + Z_{t,j}^\top \tilde b_g$, which implies  the marginal causal effect is $f_t (S_{t,j})^\top \beta$ so Assumption~\ref{ass:directeffect} holds.  Random effects in $(I)$ allow for cluster-level variation in baseline values of the proximal response, while random effects in $(II)$ allow for cluster-level variation in the fully-conditional treatment effect. Given the above generative model, sufficient condition \eqref{eq:samesiescondition} holds if $\tilde b_g \equiv 0$, i.e., when the treatment effect does not exhibit cluster-level variation.
% To see this, note that the errors $\epsilon_{t,j}$ and $\epsilon_{t,j^\prime}$ as defined in Lemma~\ref{lemma:asymnorm} do not depend on treatment if $\tilde b = 0$.  The cross-term is non-zero ($\psi \not \equiv 0$) because there is within-cluster variation due to random effects in $(I)$.
For this reason, \eqref{eq:samesiescondition} is referred to as a treatment-effect heterogeneity condition.
% i.e., when clusters exhibit treatment-effect heterogeneity this induces marginal residual correlation to depend on $a$ and $a^\prime$ which means the proposed approach is necessary for assessing direct effects rather than standard MRT analyses.
The condition motivates our simulation study in Section~\ref{section:sims}, which empirically supports this conclusion of equivalence.

\subsection{Pairwise Indirect Causal Excursion Effect Estimation}
\label{section:indirect}

\begin{assumption} \normalfont
\label{ass:indirecteffect}
Assume the pairwise indirect causal excursion effect~$\beta^{(IE)}_{{\bf p}, \pi, \Delta}(t;s) = f_t (s)^\top \beta^{\star \star}$, where $f_t (s) \in \mathbb{R}^q$ is a $q$-dimensional vector function of $s$ and time $t$.
\end{assumption}

Consider inference on the $q$-dimensional parameter $\beta^{\star \star}$. Define the weight~$W_{t,j, j^\prime}$ at decision time $t$ for the $j$th individual as equal to $\frac{\tilde p (A_{t,j}, A_{t,j^\prime} \mid S_t)}{p_t (A_{t,j}, A_{t,j^\prime} \mid H_t)}$ where $\tilde p_t (a, a^\prime \mid S_t)\in (0,1)$ is arbitrary as long as it does not depend on terms in $H_t$ other than $S_t$, and $p(A_{t,j}, A_{t,j^\prime} \mid H_t)$ is the marginal probability that individuals $j$ and $j^\prime$ receive treatments $A_{t,j}$ and $A_{t,j^\prime}$ respectively given $H_t$.  Here we consider an estimator which is the minimizer of the following cluster-based weighted-centered least-squares (C-WCLS) criterion:
\begin{align}
\mathbb{P}_M \bigg[ \frac{1}{G_m (G_m-1)} \sum_{j \neq j^\prime} \sum_{t=1}^T &W_{t,j, j^\prime} \times W_{t,\Delta, j,j^\prime} \times\bigg(Y_{t,\Delta,j} - \nonumber\\
&g_t (H_t)^\top \alpha - (1-A_{t,j}) (A_{t,j^\prime} - \tilde p_t^\star (1\mid S_t)) f_t(S_t)^\top \beta \bigg)^2 \bigg ] \label{eq:indirectwcls}
\end{align}
where $\tilde p_t^\star (1\mid S_t) = \frac{\tilde p_t (0,1 \mid S_t)}{\tilde p_t (0,0 \mid S_t) + \tilde p_t (0,1 \mid S_t)}$ and $W_{t,\Delta,j,j^\prime} = \prod_{u=t+1}^{t+\Delta-1} \pi(A_{u,j}, A_{u,j^\prime} | H_u) / p(A_{u,j}, A_{u,j^\prime} | H_u)$.  If an individual's randomization probabilities only depends on their own observed history then $\tilde p_t^\star (1 \mid S_{t,j^\prime}) = \tilde p_t (1 \mid S_{t,j^\prime})$ and the weight~$W_{t,\Delta, j,j^\prime} = W_{t,\Delta,j} \times W_{t, \Delta,j^\prime}$.  In Appendix~\ref{app:techdetails}, we prove the following result.

\begin{lemma}
\label{lemma:asymnorm2}
Under Assumption~\ref{ass:indirecteffect}, then, under invertibility and moment conditions, the estimator $\hat \beta^{(IE)}$ that minimizes \eqref{eq:indirectwcls} satisfies $\sqrt{M} \left( \hat \beta^{(IE)} - \beta^{\star \star} \right) \to N(0, Q^{-1} W Q^{-1})$ where
$$
Q = \mathbb{E} \left[ \sum_{t=1}^T (\tilde p_t (0,0 \mid S_t) + \tilde p_t (0,1 \mid S_t)) \tilde p_t^\star ( 1 \mid S_{t} ) ( 1- \tilde p_t^\star ( 1 \mid S_{t} )) f_t (S_t) f_t (S_t)^\top \right]
$$
and
\begin{align*}
W =  &\mathbb{E} \bigg[ \sum_{t=1}^T W_{t,J,J^\prime} W_{t,\Delta, J,J^\prime} \epsilon_{t,J,J^\prime} (1-A_{t,J})( A_{t,J^\prime} - \tilde p_t^\star( 1 \mid S_{t} )) f_t (S_t) \\
&\times \sum_{t=1}^T W_{t,\tilde J,\tilde J^\prime} \times W_{t, \Delta, \tilde J, \tilde J^\prime} \epsilon_{t, \tilde J, \tilde J^\prime} (1-A_{t,\tilde J}) ( A_{t,\tilde J^\prime} - \tilde p_t^\star( 1 \mid S_{t} )) f_t (S_t)^\top  \bigg]
\end{align*}
where $\epsilon_{t,j,j^\prime} = Y_{t,\Delta,j} - g_t(H_t)^\top \alpha^{\star \star} - (1-A_{t,j}) (A_{t,j^\prime} - \tilde p_t^\star (1 \mid S_t) ) f_t (S_t)^\top \beta^{\star \star}$,$\alpha^{\star \star}$ minimizes the least-squares criterion $\mathbb{E}  \left[ \frac{1}{G_m (G_m-1)} \sum_{j \neq j^\prime} \sum_{t=1}^T W_{t,j,j^\prime} W_{t,\Delta, j,j'} \left( Y_{t,\Delta,j} - g_t(H_t)^\top \alpha \right)^2 \right]$, and both $(J,J^\prime)$ and $(\tilde J, \tilde J^\prime)$ are independently, randomly sampled pairs from the cluster.
\end{lemma}

\begin{remark}%[Data-driven numerator]
In Appendix \ref{app:addindirect}, variances are re-derived in the general setting where the numerators $\tilde p_t (a \mid S_t)$ and $\tilde p_t (a, a^\prime \mid S_t)$ are estimated using the observed MRT data.
\end{remark}

\section{Simulations}
\label{section:sims}

To evaluate the proposed estimator, we extend the simulation setup in \cite{Boruvkaetal}. We first present a base data generation model, which is to be extended in four scenarios. In this section, we focus on presenting simulation settings and results for lag-1 proximal responses ($\Delta = 1$); see Appendix \ref{app:lagsimulation} for scenarios under $\Delta>1$ with similar conclusions about the relative advantage of the proposed method. Consider an MRT with known randomization probability and the observation vector $O_t$ being a single state variable $S_t \in \{-1,1\}$ at each decision time $t$. Let
\begin{equation}
\label{eq:generativemodel}
\begin{array}{r@{}l}
    Y_{t,1} = &{} \theta_1 \{S_t - \E \left[ S_t|A_{t-1},H_{t-1}\right]\} + \{A_t - p_t(1|H_t)\}(\beta_{10} + \beta_{11} S_t)+  e_{t+1}.
\end{array}
\end{equation}
The randomization probability is $p_t(1|H_t) = \text{expit}(\eta_1 A_{t-1}+\eta_2 S_t)$ where $\text{expit}(x)=(1+\exp(-x))^{-1}$; the state dynamics are given by $\mathbb{P}(S_t=1|A_{t-1},H_{t-1})=\text{expit}(\xi A_{t-1})$ with $A_0 = 0$, and the independent error term satisfies $e_t \sim \mathcal{N}(0,1)$ with $\text{Corr}(e_u, e_t) = 0.5^{|u-t|/2}$. As in~\cite{Boruvkaetal}, we set $\theta_1=0.8, \xi=0, \eta_1 = -0.8, \eta_2 = 0.8, \beta_{10}=-0.2$, and $\beta_{11} = 0.2$. Because $\xi=0$, the marginal proximal effect is equal to $\beta_{10} + \beta_{11} \E \left[S_t \right]=\beta_{10} = -0.2$.
% The marginal treatment effect is thus constant in time and is given by $\beta_1^\star = \beta_{10} =  -0.2$.
%We consider four simulation scenarios based on the above generative model. The first three concern estimation of the marginal proximal treatment effect. We set $f_t(S_t)=1$ in~\eqref{eq:generativemodel} (i.e., $S_{1t} = \emptyset$) and report average point estimate, standard deviation (SE), root mean squared error (RMSE), and 95\% confidence interval coverage probabilities (CP) across 1000 replicates. We vary the number of clusters and cluster size.  We compare the proposed cluster-based method (C-WCLS) to the approach by~\cite{Boruvkaetal} (WCLS).
In extending the data generation model to clustered settings, we conducted simulation studies with $25, 50, 100$ clusters with equal sizes ($15$ or $25$); here, we report results with $50$ clusters showing the relative advantage of C-WCLS over WCLS. A more complete set of simulation results with similar findings can be found in Appendix~\ref{app:simdetails}.

\begin{figure}
  \centering
  % \figuresize{0.8}
  % \figurebox{15pc}{20pc}{}[coveragecomparison.eps]
  \includegraphics{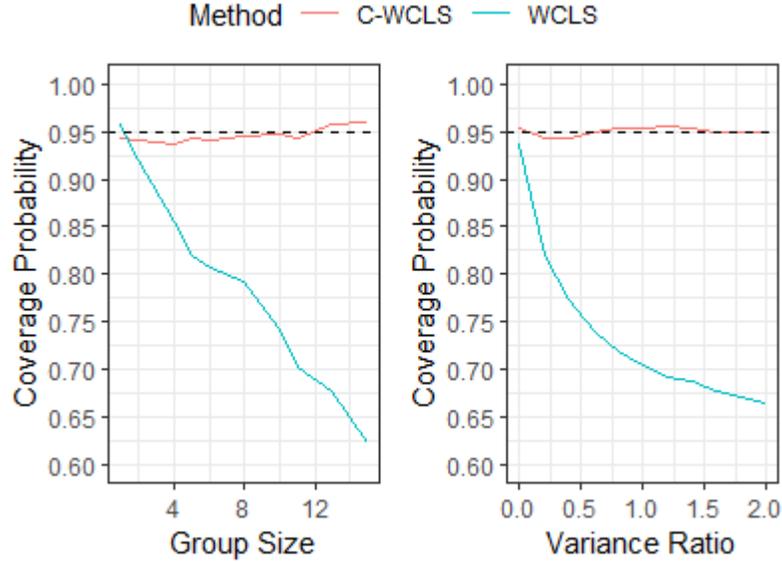}
  \caption{C-WCLS offers valid $95\%$ confidence intervals in Scenario II. WCLS does not. Empirical coverage varies by group size  ($G$) ({\bf left}), and relative variance of $b_g$ as compared to $e_g$ (\bf right).}
  \label{fig:undercoverage}
\end{figure}

% \hs{1. The 3 sims $\to$ show with no $b_g$ we recover same coverage? \zw{group-specific $b_g$ interacting or not interacting with the centered treatment. show valid coverage when $b_g$ does not interact with the centered treatments.}}

\noindent {\bf Simulation Scenario I}. The first scenario estimates the marginal proximal effect when an individual-level moderator exists and proximal responses share a random cluster-level intercept term that does not interact with treatment. The data generative model~\eqref{eq:generativemodel} incorporates a cluster-level random-intercept $e_g \sim \mathcal{N}(0,0.5)$.
% , so that $
% Y_{t+1,j} = (-0.2 + 0.2 \cdot S_{t,j}) \times (A_{t,j} -p_t(1|H_{t,j})) + 0.8 S_{t,j} + e_g +e_{t+1,j}$.
Table~\ref{tab:simresults} presents the results, which shows both WCLS and the proposed C-WCLS approach are nearly unbiased and have proper coverage rates. This is in line with Lemma~\ref{lemma:samesies} stating asymptotic equivalence under no cluster-level treatment heterogeneity.
% This demonstrates that the performance of the WCLS approach is not impacted by group-level correlation that does not interact with treatment.

\noindent {\bf Simulation Scenario II}. In the second scenario, we extend Scenario I to include a random cluster-level intercept term that interacts with treatment by considering the linear model with the additional term~$b_g \times (A_{t,j} -p_t(1|H_{t,j}))$
%\begin{equation}
%Y_{t+1,j} = (-0.2 + b_g +  0.2 \cdot S_{t,j}) \times (A_{t,j} -p_t(1|H_{t,j})) + 0.8 \cdot S_{t,j} + e_g + e_{t+1,j}
%\end{equation}
where $b_g \sim \mathcal{N}(0,0.1)$.
% is a random-intercept term within the treatment effect per cluster.
Table~\ref{tab:simresults} presents the results which demonstrate that if cluster-level random effects interact with treatment, then both methods produce nearly unbiased estimates of the marginal proximal effect but only the proposed method achieves the nominal 95\% coverage probability. To further demonstrate this, Figure~\ref{fig:undercoverage} presents nominal coverage as a function of the ratio of the variance of $b_g$ over the variance of~$e_g$ as well as group size respectively.
% This shows that the coverage probability of the WCLS method decays rapidly while the proposed method achieves the nominal 95\% coverage probability for all choices of the variance of $b_g$ and when the group size increases.
% In Figure~\ref{fig:undercoverage} (left panel), note that even when group size is $5$ (i.e., small groups), the nominal coverage drops to 80\%.

\noindent {\bf Simulation Scenario III}.  In the third scenario, the treatment effect for an individual is assumed to depend on the average state of all individuals in the cluster, i.e., define the cluster-level moderator $\bar S_{t,g} = \frac{1}{G_g}\sum_{j=1}^{G_g} S_{t,j}$ and consider the linear model from Scenario II with the additional term~$\bar S_{t,g} \times (A_{t,j} -p_t(1|H_{t,j}))$.
%\begin{equation*}
% Y_{t+1,j} = (-0.2 + b_g +  0.2 \cdot \bar S_{t,g}) \times (A_{t,j} -p_t(1|H_{t,j})) + 0.8 S_{t,j} + e_g + e_{t+1,j}
% \end{equation*}
The proposed estimator again achieves the nominal 95\% coverage probability while the WCLS method does not (see Scenario III, Table~\ref{tab:simresults}).

\begin{table}[!th]
\def~{\hphantom{0}}
\caption{Simulation: C-WCLS and WCLS comparison for Scenario I-IV.}
\begin{tabular}{c ccccccc}
\\
% \hline
Scenario & Estimator & \# of Clusters & Cluster Size & Estimate & SE & RMSE & CP \\ %\hline
\multirow{4}{*}{I}
% & C-WCLS & \multirow{2}{*}{25} & \multirow{2}{*}{10} & -0.198 & 0.035 & 0.036  & 0.945 \\
% & WCLS & & &  -0.200 & 0.036 & 0.035 & 0.956 \\  \cdashline{2-8}
% & C-WCLS & \multirow{2}{*}{25} & \multirow{2}{*}{25} & -0.199 & 0.022 & 0.023 & 0.948 \\
% & WCLS & & &  -0.199 & 0.023 & 0.022 & 0.958 \\ \cdashline{2-8}
& C-WCLS & \multirow{2}{*}{50} & \multirow{2}{*}{10} & -0.198 & 0.025 & 0.027 & 0.935 \\
& WCLS & & &  -0.198 & 0.026 & 0.026 & 0.944 \\ %\cdashline{2-8}
& C-WCLS & \multirow{2}{*}{50} & \multirow{2}{*}{25} & -0.198 & 0.016 & 0.016 & 0.950 \\
& WCLS & & &  -0.198 & 0.016 & 0.017 & 0.937 \\ %\hline  %\cdashline{2-8}
&  & & &   &  &  &  \\
% & C-WCLS & \multirow{2}{*}{100} & \multirow{2}{*}{10} & -0.199 & 0.018 & 0.018 & 0.949 \\
% & WCLS & & &    -0.198& 0.018 & 0.019 & 0.949 \\ \cdashline{2-8}
% & C-WCLS & \multirow{2}{*}{100} & \multirow{2}{*}{25} &  -0.198 & 0.011 & 0.012  & 0.941 \\
% & WCLS & & &  -0.199 & 0.011 &  0.012 & 0.944 \\ \hline
\multirow{4}{*}{II}
% & C-WCLS & \multirow{2}{*}{25} & \multirow{2}{*}{10} & -0.199 & 0.070 & 0.076 & 0.935 \\
% & WCLS & & &  -0.201 & 0.041 & 0.076 & 0.710 \\  \cdashline{2-8}
% & C-WCLS & \multirow{2}{*}{25} & \multirow{2}{*}{25} & -0.196 & 0.065 & 0.071 &   0.933 \\
% & WCLS & & &  -0.200 & 0.026 & 0.065 & 0.557 \\ \cdashline{2-8}
& C-WCLS & \multirow{2}{*}{50} & \multirow{2}{*}{10} & -0.200 & 0.051 & 0.049 & 0.957 \\
& WCLS & & &  -0.200 & 0.029 & 0.052 & 0.723 \\ %\cdashline{2-8}
& C-WCLS & \multirow{2}{*}{50} & \multirow{2}{*}{25} & -0.200 & 0.047 & 0.049 & 0.947 \\
& WCLS & & &  -0.199 & 0.019 & 0.048 & 0.555 \\ %\hline %\cdashline{2-8}
&  & & &   &  &  &  \\
% & C-WCLS & \multirow{2}{*}{100} & \multirow{2}{*}{10} & -0.198  & 0.036 & 0.035 & 0.955 \\
% & WCLS & & &  -0.198 & 0.021 & 0.037 & 0.718 \\ \cdashline{2-8}
% & C-WCLS & \multirow{2}{*}{100} & \multirow{2}{*}{25} & -0.198  & 0.033 & 0.035 & 0.942 \\
%& WCLS & & &  -0.199 & 0.013 & 0.032 & 0.583 \\ \hline
\multirow{4}{*}{III}
% & C-WCLS & \multirow{2}{*}{25} & \multirow{2}{*}{10} & -0.199 & 0.070 & 0.073 & 0.948 \\
% & WCLS & & &  -0.202 & 0.041 & 0.074 & 0.734 \\  \cdashline{2-8}
% & C-WCLS & \multirow{2}{*}{25} & \multirow{2}{*}{25} & -0.196 & 0.066 & 0.069 & 0.931 \\
% & WCLS & & &  -0.200 & 0.026 & 0.068 & 0.563 \\ \cdashline{2-8}
& C-WCLS & \multirow{2}{*}{50} & \multirow{2}{*}{10} & -0.198 & 0.051 & 0.052 & 0.941 \\
& WCLS & & &  -0.199 &  0.029 & 0.052 & 0.742 \\ %\cdashline{2-8}
& C-WCLS & \multirow{2}{*}{50} & \multirow{2}{*}{25} & -0.199 & 0.047 & 0.048 & 0.946 \\
& WCLS & & &  -0.200 & 0.018 & 0.048 & 0.561 \\% \hline %\cdashline{2-8}
&  & & &   &  &  &  \\
% & C-WCLS & \multirow{2}{*}{100} & \multirow{2}{*}{10} & -0.200 & 0.036 & 0.037 & 0.946 \\
% & WCLS & & &    -0.200 & 0.021 & 0.037 & 0.740 \\ \cdashline{2-8}
% & C-WCLS & \multirow{2}{*}{100} & \multirow{2}{*}{25} & -0.201 & 0.034 & 0.033 & 0.956 \\
% & WCLS & & &  -0.198 & 0.013 & 0.034 & 0.555 \\ \hline
\multirow{2}{*}{IV}
& \multirow{2}{*}{C-WCLS} & \multirow{2}{*}{50} &
10 & -0.097 & 0.020 & 0.021 & 0.953 \\
& & & 25 & -0.100 & 0.013 & 0.013 & 0.942 \\ %\hline
\end{tabular}
\label{tab:simresults}
%\begin{tabnote}
%U.S., United States of America; R, respondent.
%\end{tabnote}
\end{table}

% \hs{4. Demonstrate indirect effect}

\noindent {\bf Simulation Scenario IV}.  The fourth scenario considers the indirect effect.  For individual $j$ at decision point $t$, define the total effect to be $TE_{t,j} = \sum_{j^\prime \neq j} \{A_{t,j^\prime} - \tilde p_{t, j^\prime} ( 1 \mid H_t) \} (\beta_{20} + \beta_{21} S_{t,j^\prime})$, where $\beta_{20} = -0.1$ and $\beta_{21} = 0.2$. The generative model is then given by:
\begin{equation*}
    Y_{t,1,j} = (-0.2 + b_g +  0.2 \cdot \bar S_{t,g}) \times \{A_{t,j} -p_t(1|H_{t,j})\}+ 0.8 S_{t,j} +TE_{t,j} +e_g +e_{t+1,j}.
\end{equation*}
This model implies a marginal pairwise indirect effect equal to $\beta^{(IE)} = \beta_{20} = -0.1$. Table~\ref{tab:simresults} presents simulation results which shows that the proposed indirect estimator exhibited nearly no bias and achieved the nominal coverage probability.

% \noindinet \begin{remark}
% A simulation study of direct and pairwise indirect effects with $\Delta>1$ is presented in Appendix \ref{app:lagsimulation}.
% \end{remark}
% \begin{table}[!th]
% \def~{\hphantom{0}}
% \tbl{\it Simulations show strong finite sample estimation and accurate coverage for indirect effects.}{%
% \begin{tabular}{c | cccccc}
% \hline
% Scenario & \# of Clusters & Cluster Size & Estimate & SD & RMSE & CP \\ \hline
% \multirow{6}{*}{IV} & 25 & 10 & -0.097 & 0.028 & 0.029 & 0.958 \\
% & 25 & 25 & -0.101 & 0.017 &  0.019 & 0.942 \\
% & 50 & 10 & -0.097 & 0.020 & 0.021 & 0.953 \\
% & 50 & 25 & -0.100 & 0.013 & 0.013 & 0.942 \\
% & 100 & 10 & -0.097 & 0.015 & 0.015 & 0.943 \\
% & 100 & 25 &  -0.100 & 0.009 & 0.009 & 0.944 \\ \hline
% \end{tabular}}
% \label{tab:simresults_indirect}
% \end{table}

\section{Case Study: Intern Health Study}
\label{section:casestudy}

The Intern Health Study (IHS) was a 6-month MRT on 1,562 medical interns where four types of weekly notification - mood, activity, sleep, or none -- were randomly assigned with equal probability to each subject~\citep{Necamp2020}; see Section~\ref{section:motex} for prior discussion.
In IHS,  285 institutions and 24 specialties were observed.   Here, we assess the effect of the three types of notifications (mood, activity, and sleep) compared to no notifications on the weekly average of self-reported mood scores, log step-count and log sleep minutes for the population of interns.
% See Appendix~\ref{app:IHSadditionalanalysis} for an additional analysis of log sleep minutes.
Due to high levels of missing data, weekly proximal responses were multiply imputed.
% Similar imputation was performed for daily log step-counts and daily log sleep minutes.
See~\cite{Necamp2020} for further details.

\begin{figure}[!th]
  % \figuresize{0.8}
  % \figurebox{15pc}{20pc}{}[mainpaper.eps]
  \includegraphics{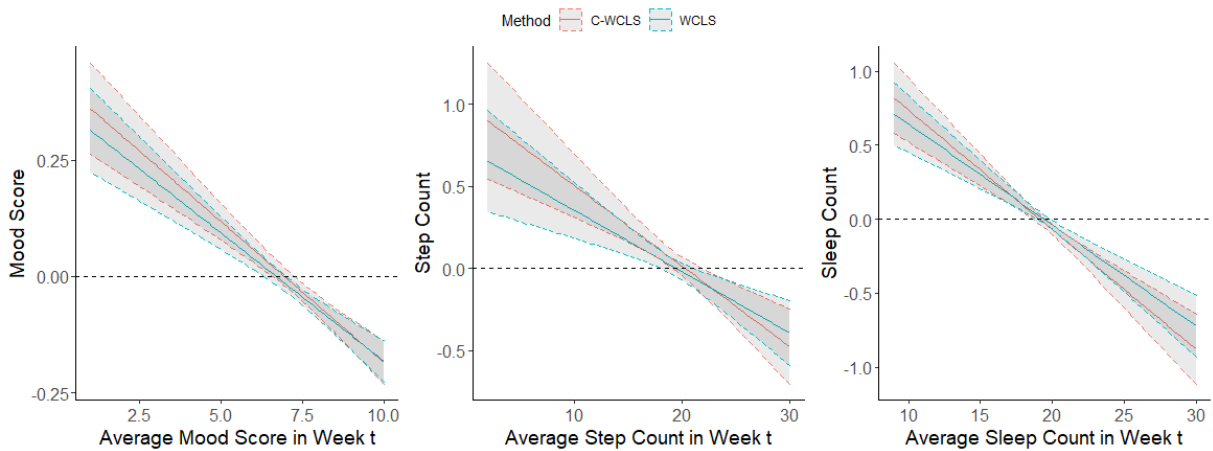}
  \caption{Moderation of average previous week's proximal responses on the effect of notifications on average weekly mood scores, log step counts, and log sleep counts respectively in IHS. }
    \label{fig:wcls_moderation_IHS}
\end{figure}

Let $t=1,\ldots,T$ denote the weekly decision points at which the individual is randomized to the various types of notifications.
% Because of the form of the intervention, all participants were available for this intervention throughout the study; i.e., $I_t \equiv 1$.
The three proximal responses are the average weekly mood score,  which is reported on a Likert scale  taking values from 1 to 10 (higher scores mean better mood), log step count and log sleep minutes respectively.
Notifications are collapsed to a binary variable, i.e., $A_{t,j}=1$ if the individual was assigned to receive any notifications on week $t$; otherwise, $A_{t,j}= 0$.
% At any occasion $t$, an individual's notification randomization probabilities were only dependent of their past observed history~$H_{t,j}$.
% For simplicity,
We start by defining clusters based on medical specialty as we saw effect heterogeneity by specialty in Figure~\ref{fig:wcls_heterogeneity}.  The average cluster size was 65; the first and third quartile were 7 and 113 respectively, with maximum and minimum sizes of 333 and 1.  For every individual in each cluster at each decision point, we compute the average prior weekly proximal response for all others in the cluster, denoted $\bar Y_{t, -j}$ for the $j$th individual in the cluster.  We conducted analyses under lag $\Delta=1$ and $\Delta=2$. Here we report results under $\Delta=1$; see Appendix~\ref{app:IHSadditionalanalysis} for results under $\Delta=2$ for two choices of reference policy $\pi$. Under $\Delta=1$, we consider two moderation analyses that can both be expressed as
$\beta(t; S_t) = \beta_0 + \beta_1 \cdot Y_{t,j} + \beta_2 \bar Y_{t,-j}$.

The first set of moderation analyses considers the standard moderation analysis where only individual-level moderators are included (i.e., $\beta_2 = 0$). Figure~\ref{fig:wcls_moderation_IHS} visualizes the estimates across the range of prior week's proximal response for both our proposed approach and the WCLS approach from~\cite{Boruvkaetal}, and the numerical output can be found in Appendix ~\ref{app:moreonFigure3}. In comparison, C-WCLS produces larger variance estimates for all proximal responses as expected. The effects do not change too much for the average weekly mood and sleep analysis; however, the significant effect of messages on weekly log step count under the traditional MRT analysis becomes insignificant when accounting for cluster effects.

The second moderation analysis lets $\beta_2$ be a free parameter, enabling novel moderation analyses that accounts for the average weekly previous proximal responses of other individuals.  Table~\ref{tab:IHS_direct} presents the results.  Here, we see that the new term $\beta_2$ is negative but insignificant.  The results suggest the average proximal responses of others in the cluster have a limited moderation effect. To conclude, the impact of a notification on mood is larger while the individual's score from previous week is low.   Similar results hold for the log step-count analysis.
% the constant term~$\beta_0$  while

% \begin{table}[!th]
% \def~{\hphantom{0}}
% \tbl{\it Moderation analysis with cluster-level moderators.}{%
% \begin{tabular}{c |c | crrrr}
% \hline
% Outcome & Setting & Variables & Estimate & Std. Error & t-value & p-value \\ \hline
% \multirow{3}{*}{Mood}
% % & \multirow{2}{*}{WCLS} & $\beta_0$ & 0.369 & 0.086 & 4.268 & 0.000 \\
% % & & $\beta_1$ & -0.055 & 0.011 & -4.822 & 0.000 \\ \cline{2-7}
% % & \multirow{2}{*}{C-WCLS} & $\beta_0$ & 0.350 & 0.103 & 3.401 & 0.001 \\
% % & & $\beta_1$ & -0.053 & 0.014 & -3.868 & 0.000 \\ \cline{2-7}
% & \multirow{3}{*}{C-WCLS} & $\beta_0$ & -0.238 & 0.282 & -0.842 & 0.401 \\
% & & $\beta_1$ & -0.054 & 0.014 & -3.973 & 0.000 \\
% & & $\beta_2$ & 0.083 & 0.037 & 2.241 & 0.026 \\ \hline
% \multirow{3}{*}{Steps}
% % & \multirow{2}{*}{WCLS} & $\beta_0$ & 0.729 & 0.295 & 2.472 & 0.015 \\
% % & & $\beta_1$ & -0.037 & 0.015 & -2.484 & 0.015 \\ \cline{2-7}
% % & \multirow{2}{*}{C-WCLS} & $\beta_0$ & 0.622 & 0.384 & 1.618 & 0.108 \\
% % & & $\beta_1$ & -0.031 & 0.019 & -1.580 & 0.117 \\ \cline{2-7}
% & \multirow{3}{*}{C-WCLS} & $\beta_0$ & -2.095 & 1.248 & -1.678 & 0.094 \\
% & & $\beta_1$ & -0.034 & 0.019 & -1.745 & 0.083 \\
% & & $\beta_2$ & 0.143 & 0.062 &2.330 & 0.020 \\ \hline
% \multirow{3}{*}{Sleep}
% & \multirow{3}{*}{C-WCLS} & $\beta_0$ & -1.912 & 1.379 & -1.386 & 0.171  \\
% & &$\beta_1$ & -0.067 & 0.023 & -2.948 & 0.004 \\
% & &$\beta_2$ & 0.162 & 0.065 & 2.487 & 0.015 \\ \hline
% \end{tabular}}
% \label{tab:IHS_direct}
% \end{table}

\begin{table}
{\scriptsize
% \caption{Moderation analysis using C-WCLS in IHS (lag $\Delta=1$)}
\begin{tabular}{lcrrrcrrr}
\\
& \multicolumn{4}{c}{Direct Effect} & \multicolumn{4}{c}{Indirect Effect} \\
& Variables & Estimate & Std. Error & p-value & Variables & Estimate & Std. Error & p-value \\[5pt]
\multirow{3}{*}{Mood}
  & Intercept ($\beta_0$) & 0.563 & 0.251  & 0.028  & $\tilde \beta_0$ & -0.054 & 0.045  & 0.883 \\
 & Prior Week Avg. ($\beta_1$) & -0.066  &   0.027    & 0.016 & $\tilde \beta_1$ & -0.015 & 0.031  & 0.684 \\
 & Cluster Pr. Wk. Avg. ($\beta_2$) & -0.016  &   0.017  &   0.349 & & & & \\
%  &&  & &   & & & & &  \\
\multirow{3}{*}{Steps}
 & Intercept ($\beta_0$) & 1.165  & 0.782   &  0.139 & $\tilde \beta_0$ & -0.038 & 0.134  & 0.612 \\
 & Prior Week Avg. ($\beta_1$) & -0.048  & 0.036 &  0.177 & $\tilde \beta_1$ & 0.019 & 0.090  & 0.417\\
 & Cluster Pr. Wk. Avg. ($\beta_2$) & -0.010  &   0.014    & 0.482 & & & & \\ %\hline
% &&  & &   &  & & & & \\
\multirow{3}{*}{Sleep}
 & Intercept ($\beta_0$) & 1.545 &  0.779  &  0.050 & $\tilde \beta_0$ & 0.007 & 0.096 & 0.469 \\
 & Prior Week Avg. ($\beta_1$) & -0.081   &  0.039  &    0.037 & $\tilde \beta_1$ & -0.004 & 0.065 &  0.526 \\
 & Cluster Pr. Wk. Avg. ($\beta_2$) & 0.000 & 0.006   & 0.961 & & & &\\
\end{tabular}
}
\caption{
Moderation analysis for the direct and indirect effect of notifications on average weekly mood scores, log step counts, and log sleep minutes respectively in IHS.  Coefficient $\tilde \beta_0$ represents the indirect effect under $A_{t,j} = 0$, while the coefficient $\tilde \beta_1$ represents the indirect effect under $A_{t,j} = 1$.
}
\label{tab:IHS_direct}
\end{table}

% \begin{table}[!th]
% \def~{\hphantom{0}}
% \tbl{\it Moderation analysis with cluster-level moderators (lag $\Delta=1$).}{%
% \begin{tabular}{c c crrrr}
% %\hline
% Outcome &  Variables & Estimate & Std. Error & p-value \\ %\hline
% \multirow{3}{*}{Mood}
% % & \multirow{2}{*}{WCLS} & $\beta_0$ & 0.369 & 0.086 & 4.268 & 0.000 \\
% % & & $\beta_1$ & -0.055 & 0.011 & -4.822 & 0.000 \\ \cline{2-7}
% % & \multirow{2}{*}{C-WCLS} & $\beta_0$ & 0.350 & 0.103 & 3.401 & 0.001 \\
% % & & $\beta_1$ & -0.053 & 0.014 & -3.868 & 0.000 \\ \cline{2-7}
%   & $\beta_0$ & 0.563 & 0.251  & 0.028 \\
%  & $\beta_1$ & -0.066  &   0.027    & 0.016 \\
%  & $\beta_2$ & -0.016  &   0.017  &   0.349 \\ %\hline
%  &&  & &   &  \\
% \multirow{3}{*}{Steps}
% % & \multirow{2}{*}{WCLS} & $\beta_0$ & 0.729 & 0.295 & 2.472 & 0.015 \\
% % & & $\beta_1$ & -0.037 & 0.015 & -2.484 & 0.015 \\ \cline{2-7}
% % & \multirow{2}{*}{C-WCLS} & $\beta_0$ & 0.622 & 0.384 & 1.618 & 0.108 \\
% % & & $\beta_1$ & -0.031 & 0.019 & -1.580 & 0.117 \\ \cline{2-7}
%   & $\beta_0$ & 1.165  & 0.782   &  0.139 \\
%  & $\beta_1$ & -0.048  & 0.036 &  0.177 \\
%  & $\beta_2$ & -0.010  &   0.014    & 0.482 \\ %\hline
%  &&  & &   &  \\
% \multirow{3}{*}{Sleep}
%   & $\beta_0$ & 1.545 &  0.779  &  0.050  \\
%  &$\beta_1$ & -0.081   &  0.039  &    0.037 \\
%  &$\beta_2$ & 0.000 & 0.006   & 0.961 \\ %\hline
% \end{tabular}}
% \label{tab:IHS_direct}
% \end{table}

Finally, we consider indirect moderation effect analyses.   In this analysis, clusters are defined based on medical specialty and institution because interference was only likely when interns are in close geographic proximity.   Here, we consider the marginal indirect effect (e.g., no moderators) both when the individual did not receive the intervention and when the individual did receive an intervention at decision time~$t$.  Table~\ref{tab:IHS_direct} presents the results.  In this case, the estimated indirect effects are much weaker than the direct effects.  Even a weak effect may be unexpected as none of the content in the push notifications was aimed at impacting other individuals' behavior.  For all the proximal responses, we see limited evidence of an indirect effect.  This implies that the scientific team, when building an optimal intervention package, may ignore these indirect effects and focus solely on the individual who receives these types of push notifications.
% Existence of signal for the impact on proximal mood implies the scientific team may wish to consider interventions that target these indirect effects more explicitly, either in the framing of the messages or in messages aimed at impacting the overall cluster.
% \begin{table}[!th]
% \def~{\hphantom{0}}
% \tbl{\it Moderation analysis for the indirect effect of notifications on average weekly mood scores, log step counts, and log sleep minutes respectively in IHS.  Coefficient $\tilde \beta_0$ represents the indirect effect under $A_{t,j} = 0$, while the coefficient $\tilde \beta_1$ represents the indirect effect under $A_{t,j} = 1$.}{%
% \begin{tabular}{ccrrrr}
% %\hline
% Outcome & Variables & Estimate & Std. Error  & p-value \\ %\hline
% \multirow{2}{*}{Mood} & $\tilde \beta_0$ & -0.054 & 0.045  & 0.883 \\
% & $\tilde \beta_1$ & -0.015 & 0.031  & 0.684 \\ %\hline
% & &  & & & \\
% \multirow{2}{*}{Steps} & $\tilde \beta_0$ & -0.038 & 0.134  & 0.612\\
% & $\tilde \beta_1$ & 0.019 & 0.090  & 0.417 \\ %\hline
% & &  & & & \\
% \multirow{2}{*}{Sleep} & $\tilde \beta_0$ & 0.007 & 0.096 & 0.469\\
% & $\tilde \beta_1$ & -0.004 & 0.065 &  0.526\\ %\hline
% \end{tabular}}
% \label{tab:IHS_indirect}
% \end{table}

\section{Discussion}

We revisited causal excursion effects in the presence of \textit{a priori} known clusters in sequential treatment settings where outcome of interest is measured at all decision points. In particular, we formalized both direct and indirect excursion effects in the context of MRT to account for potential interference. We studied their identifications and proposed consistent and robust inference methods.  In practice, the effects described in this paper are most important when using MRT data to build optimized just-in-time adaptive interventions (JITAIs) for deployment in an mHealth package. Specifically, the estimation procedure for the direct excursion effect accounts for within-cluster correlation in the proximal responses which helps the scientific team avoid making erroneous conclusions about intervention effectiveness using standard MRT methods.  Moreover, estimation of indirect effects allows the scientific team to answer questions about impact of interventions on other members of the same cluster.  Use of these methods provides empirical evidence for the scientific team to include or exclude intervention components that may have had unanticipated second order effects, or potentially lead to novel ways to improve the intervention component by revising the intervention to more explicitly account for cluster-level interference. While this work represents a major step forward in the analysis of micro-randomized trial data, further work is required.  Specifically, useful extensions include accounting for overlapping communities and/or network (rather than cluster-only) structure~\citep{Ogburn2014,Mealli2019}, accounting for general non-continuous proximal responses such as binary or count outcomes~\citep{Qian2021}, penalization of the working model to allow for high-dimensional moderators, and a method to use the proposed approach to form warm-start policies at the individual level while accounting for group level information~\citep{Luckett2020}.

% \section*{Acknowledgments}
% The authors would like to thank the Intern Health Study (IHS) team at University of Michigan for substantive discussions, Professor Srijan Sen (Principal Investigator) for generously providing access to the IHS data, and the study participants of IHS for providing the data.  The draft also benefited from helpful comments from Professors Inbal Nahum-Shani and Peng Ding.

\bibliographystyle{unsrt}
\bibliography{paper-ref}

\newcommand{\noop}[1]{}
\begin{thebibliography}{10}

\bibitem{Nahum2017}
I.~Nahum-Shani, S.N. Smith, B.J. Spring, L.M. Collins, K.~Witkiewitz,
  A.~Tewari, and S.A. Murphy.
\newblock Just-in-time adaptive interventions ({JITAI}s) in mobile health: Key
  components and design principles for ongoing health behavior support.
\newblock {\em Annals of Behavioral Medicine}, 2017.

\bibitem{KlasnjaMRT}
P.~Klasnja, E.B. Hekler, S.~Shiffman, A.~Boruvka, D.~Almirall, A.~Tewari, and
  S.A. Murphy.
\newblock Microrandomized trials: An experimental design for developing
  just-in-time adaptive interventions.
\newblock {\em Health Psychol}, 34:1220--1228, 2015.

\bibitem{Boruvkaetal}
Audrey Boruvka, Daniel Almirall, Katie Witkiewitz, and Susan~A. Murphy.
\newblock Assessing time-varying causal effect moderation in mobile health.
\newblock {\em Journal of the American Statistical Association},
  113(523):1112--1121, 2018.

\bibitem{Qian2021}
Tianchen Qian, Hyesun Yoo, Predrag Klasnja, Daniel Almirall, and Susan Murphy.
\newblock {Estimating time-varying causal excursion effect in mobile health
  with binary outcomes}.
\newblock {\em Biometrika}, 09 2020.
\newblock asaa070.

\bibitem{DempseyAOAS}
W.~Dempsey, P.~Liao, and S.A. Murphy.
\newblock The stratified micro-randomized trial design: Sample size
  considerations for testing nested causal effects of time-varying treatments.
\newblock {\em Annals of Applied Statistics}, 14(2):661--684, 2020.

\bibitem{Liaoetal2015}
P.~Liao, P.~Klasjna, A.~Tewari, and S.A. Murphy.
\newblock Micro-randomized trials in mhealth.
\newblock {\em Statistics in Medicine}, 35(12):1944--71, 2016.

\bibitem{Rubin}
DB. Rubin.
\newblock Bayesian inference for causal effects: The role of randomization.
\newblock {\em The Annals of Statistics}, 6(1):34--58, 1978.

\bibitem{Robins}
J.~Robins.
\newblock A new approach to causal inference in mortality studies with a
  sustained exposure period-application to control of the healthy worker
  survivor effect.
\newblock {\em Mathematical Modelling}, 7(9):1393--1512, 1986.

\bibitem{Liang1986}
KY~Liang and SL~Zeger.
\newblock Longitudinal data analysis using generalized linear models.
\newblock {\em Biometrika}, 73(1):13--22, 1986.

\bibitem{Necamp2020}
Timothy NeCamp, Srijan Sen, Elena Frank, Maureen~A Walton, Edward~L Ionides,
  Yu~Fang, Ambuj Tewari, and Zhenke Wu.
\newblock Assessing real-time moderation for developing adaptive mobile health
  interventions for medical interns: Micro-randomized trial.
\newblock {\em Journal of medical Internet research}, 22(3):e15033, March 2020.

\bibitem{Halloran1995}
ME~Halloran and CJ~Struchiner.
\newblock Causal inference for infectious diseases.
\newblock {\em Epidemiology}, 6:142--151, 1995.

\bibitem{Tchetgen2012}
E.~J. Tchetgen~Tchetgen, M.M. Glymour, J.~Weuve, and Robins J.
\newblock Specifying the correlation structure in inverse-probability-weighting
  estimation for repeated measures.
\newblock {\em Epidemiology}, 23(4):644--646, 2012.

\bibitem{Pearl2009}
J.~Pearl.
\newblock Causal inference in statistics: An overview.
\newblock {\em Statistics Surveys}, 3:96--146, 2009.

\bibitem{MurphyJASA2001}
S~A Murphy, M~J van~der Laan, J~M Robins, and Conduct Problems
  Prevention~Research Group.
\newblock Marginal mean models for dynamic regimes.
\newblock {\em Journal of the American Statistical Association},
  96(456):1410--1423, 2001.
\newblock PMID: 20019887.

\bibitem{Collins2018}
L.M. Collins.
\newblock {\em Optimization of Behavioral, Biobehavioral, and Biomedical
  Interventions}.
\newblock Springer International Publishing, 2018.

\bibitem{Hong2006}
G.~Hong and S.~W. Raudenbush.
\newblock Evaluating kindergarten retention policy.
\newblock {\em Journal of the American Statistical Association},
  101(475):901--910, 2006.

\bibitem{Vanderweele2013}
T.~J. Vanderweele, G.~Hong, S.M. Jones, and J.L. Brown.
\newblock Mediation and spillover effects in group-randomized trials: A case
  study of the 4{R}s educational intervention.
\newblock {\em Journal of the American Statistical Association},
  108(502):469--482, 2013.

\bibitem{Neugebauer2007}
R.~Neugebauer and M.~J. van~der Laan.
\newblock Nonparametric causal effects based on marginal structural models.
\newblock {\em Journal of Statistical Planning and Inference}, 137:419--434,
  2007.

\bibitem{Rosenblum2010}
M.~A. Rosenblum and M.~J. van~der Laan.
\newblock Targeted maximum likelihood estimation of the parameter of a marginal
  structural model.
\newblock {\em The International Journal of Biostatistics}, 6, 2010.

\bibitem{Kennedy2019}
Edward~H. Kennedy, Scott Lorch, and Dylan~S. Small.
\newblock Robust causal inference with continuous instruments using the local
  instrumental variable curve.
\newblock {\em Journal of the Royal Statistical Society: Series B (Statistical
  Methodology)}, 81(1):121--143, 2019.

\bibitem{Ogburn2014}
E.~Ogburn and T.~VanderWeele.
\newblock Causal diagrams for interference.
\newblock {\em Statistical Science}, 29(4):559--578, 2014.

\bibitem{Mealli2019}
Georgia Papadogeorgou, Fabrizia Mealli, and Corwin~M. Zigler.
\newblock Causal inference with interfering units for cluster and population
  level treatment allocation programs.
\newblock {\em Biometrics}, 75(3):778--787, 2019.

\bibitem{Luckett2020}
Daniel~J. Luckett, Eric~B. Laber, Anna~R. Kahkoska, David~M. Maahs, Elizabeth
  Mayer-Davis, and Michael~R. Kosorok.
\newblock Estimating dynamic treatment regimes in mobile health using
  v-learning.
\newblock {\em Journal of the American Statistical Association},
  115(530):692--706, 2020.
\newblock PMID: 32952236.

\bibitem{Mancl2001}
LA. Mancl and T.A. DeRouen.
\newblock A covariance estimator for {GEE} with improved small-sample
  properties.
\newblock {\em Biometrics}, 57(1):126--134, 2001.

\bibitem{Robins1994}
J.~M. Robins.
\newblock Correcting for non-compliance in randomized trials using structural
  nested mean models.
\newblock {\em Communications in Statistics-Theory and Methods}, 23:2379--2412,
  1994.

\bibitem{Newey1990}
W.K. Newey.
\newblock Semiparametric efficiency bounds.
\newblock {\em Journal of Applied Econometrics}, (5):99--135, 1990.

\bibitem{Tsiatis2007}
A.~Tsiatis.
\newblock {\em Semiparametric Theory and Missing Data}.
\newblock Springer Science \& Business Media, 2007.

\end{thebibliography}

\appendix

\section{Technical Details}
\label{app:techdetails}

\begin{proof}[Lemma \ref{lemma:cond_effect}]
We establish Lemma~\ref{lemma:cond_effect} for the direct effect~\eqref{eq:directavglineareffect}. For~$a_s \in \{ 0,1\}^{G}$, we consider
\begin{align*}
\E \bigg[&\left(\prod_{s=1}^{t-1} p_s(a_s |H_s(\bar a_{s-1})) \right) \left( \prod_{j^\prime \neq j} \prod_{s=t}^{t+\Delta-1} p_s(a_{s,j^\prime} \mid H_s(\bar a_{s-1})) \right)  \\
&\left( \prod_{s=t}^{t+\Delta-1} \pi_s(a_{s,j} \mid H_s(\bar a_{s-1})) \right)
 Y_{t,\Delta,j} (\bar{a}_{t+\Delta-1,-j}, (\bar a_{t-1,j},a, a_{t+1:(t+\Delta-1), j})) {\bf 1}_{S_t(\bar a_t) = s} \bigg] \\
= \E \bigg[&\left(\prod_{s=1}^{t-1} p_s(a_s |H_s(\bar a_{s-1})) \right) {\bf 1}_{S_t(\bar a_t) = s} \left( \prod_{j^\prime \neq j} \prod_{s=t}^{t+\Delta-1} p_s(a_{s,j^\prime} \mid H_s(\bar a_{s-1})) \right)  \\
&\E \left[ \left( \prod_{s=t}^{t+\Delta-1} \pi_s(a_{s,j} \mid H_s(\bar a_{s-1})) \right) Y_{t,\Delta,j} (\bar{a}_{t+\Delta-1,-j}, (\bar a_{t-1,j},a, a_{t+1:(t+\Delta-1), j})) \mid H_t (\bar a_{t-1}) \right] \bigg] \\
= \E \bigg[&\left(\prod_{s=1}^{t-1} p_s(a_s |H_s(\bar a_{s-1})) \right) {\bf 1}_{S_t(\bar a_t) = s} \left( \prod_{j^\prime \neq j} \prod_{s=t}^{t+\Delta-1} p_s(a_{s,j^\prime} \mid H_s(\bar a_{s-1})) \right)  \\
&\E \bigg[ \left( W_{t,\Delta, j} (\bar a_{t+\Delta-1}) \prod_{s=t}^{t+\Delta-1} p_s(a_s |H_s(\bar a_{s-1})) \right) Y_{t,\Delta,j} (\bar{a}_{t+\Delta-1,-j}, (\bar a_{t-1,j},a, a_{t+1:(t+\Delta-1), j})) \mid H_t (\bar a_{t-1}) \bigg] \bigg] \\
\end{align*}
since the history $H_t$ includes the moderator variable $S_t$ at time $t$. By consistency,~$H_t(\bar{A}_{t-1}) = H_t$.
%  \begin{align*}
%  %\label{eq:cons_version}
% \E \bigg[&\left(\prod_{s=1}^{t-1} p_s(a_s |H_s(\bar a_{s-1})) \right) {\bf 1}_{S_t(\bar a_t) = s} \left( \prod_{j^\prime \neq j} \prod_{s=t}^{t+\Delta-1} p_s(a_{s,j^\prime} \mid H_s(\bar a_{s-1})) \right)  \\
% &\times \E \left[ \left( \prod_{s=t}^{t+\Delta-1} \pi_s(a_{s,j} \mid H_s(\bar a_{s-1})) \right) Y_{t,\Delta,j} (\bar{a}_{t+\Delta-1,-j}, (\bar a_{t-1,j},a, a_{t+1:(t+\Delta-1), j})) \mid H_t \right] \bigg] \\
%  \end{align*}
Moreover, sequential ignorability implies that
\begin{align*}
 \E &\left[ \left( W_{t,\Delta, j} (\bar a_{t+\Delta-1}) \prod_{s=t}^{t+\Delta-1} p_s(a_s |H_s(\bar a_{s-1})) \right) Y_{t,\Delta,j} (\bar{a}_{t+\Delta-1,-j}, (\bar a_{t-1,j},a, a_{t+1:(t+\Delta-1),j})) \given H_t \right]  \\
 = \E &\left[ \left( W_{t,\Delta, j} (\bar a_{t+\Delta-1}) \prod_{s=t}^{t+\Delta-1} p_s(a_s |H_s(\bar a_{s-1})) \right) Y_{t,\Delta,j} (\bar{a}_{t+\Delta-1,-j}, (\bar a_{t-1,j},a, a_{t+1:(t+\Delta-1),j})) \given H_{t}, A_{t,j} = a \right]
\end{align*}
It also implies that $\E \left[ 1[A_{t+k,j} = a] \mid H_{t+k} \right] \cdot \E \left[ Y_{t,\Delta,j} \mid H_{t+k} \right] = \E \left[ Y_{t,\Delta,j} 1[A_{t+k,j} = a] \mid H_{t+k} \right]$.
Summing over all potential outcomes yields
\begin{align*}
\E &\bigg [ \sum_{\bar{a}_{t-1}, \bar a_{t+1:(t+\Delta-1)}} \left(\prod_{s=1}^{t-1} p_s(a_s|H_s) \right) \left( \prod_{j^\prime \neq j} p_t(a_{t,j^\prime} \mid H_t \right)  {\bf 1}_{S_t(\bar a_t) = s} \\
&\times \E \left[ \left( \prod_{s=t}^{t+\Delta-1} \pi_s(a_{s,j} \mid H_t \right) Y_{t,\Delta,j} (\bar{a}_{t+\Delta-1,-j}, (\bar a_{t-1,j},a, a_{t+1:(t+\Delta-1),j})) \given H_t, A_{t,j} = a \right] \, \bigg] \\
 = \E &\bigg [ \sum_{\bar{a}_{t-1}, \bar a_{t+1:(t+\Delta-1)}} \left(\prod_{s=1}^{t-1} p_s(a_s|H_s) \right) \left( \prod_{j^\prime \neq j} p_t(a_{t,j^\prime} \mid H_t \right)  {\bf 1}_{S_t = s} \\
 &\times
 \E \bigg[ \left( \prod_{s=t}^{t+\Delta-1} p_s(a_{s} \mid H_t) \right) \times W_{t,\Delta,j} (\bar a_{t+\Delta-1}) \\
 &\times Y_{t,\Delta,j} (\bar{a}_{t+\Delta-1,-j}, (\bar a_{t-1,j},a, a_{t+1:(t+\Delta-1),j})) \given H_t, A_{t,j} = a \bigg]
 \given S_t = s \bigg] \\
 \Rightarrow
 \E_{{\bf p}, \pi} &\left[ Y_{t,\Delta,j} (\bar A_{t+\Delta-1, -j}, (\bar A_{t-1,j}, a, \tilde A_{t+1:t+\Delta-1}) \mid S_t (\bar A_{t-1}) = s \right] =
 \E \left[ \E \left[ W_{t,\Delta,j} Y_{t,\Delta,j}  \given H_t, A_t = a \right] \given S_t = s \right].
\end{align*}
% In the final equation, the outer expectation is with respect to the history~$H_t$ conditional on~$S_t=s$.  That is, over \emph{both} past treatments~$A_s$
% and past observations~$O_s$ for $s < t$ as well as over current treatments for $A_{t,j^\prime}$ for $j^\prime \neq j$. The above shows
% $$
% \E \left[ Y_{t,1} (\bar{A}_{t,-j}, (\bar A_{t-1,j},a)) \mid S_t (\bar a_t) = s \right] = \E \left[ \E \left[Y_{t+1, j}  \given H_t, A_{t,j} = a \right] \given S_t = s \right].
% $$
Averaging over individuals in the group $j \in [G]$ group size completes the proof.
The proof for the indirect effect follows the exact same structure.
\end{proof}

\section{Lemma~\ref{lemma:asymnorm}}
\label{app:asymptotics}

We next provide a detailed proof of asymptotic normality and consistency
for the weighted-centered least squares estimator.

\begin{proof}[Consistency for direct and indirect effects]
The solutions~$(\hat \alpha,\hat \beta)$ that minimize equation~\eqref{eq:directwcls} are consistent estimators for the
solutions that minimize the following
\[
\E \left[ \frac{1}{G} \sum_{j=1}^G \sum_{t=1}^T W_{t,j} \times W_{t,\Delta,j} \left( Y_{t,\Delta,j} - g_t(H_t)^\top \alpha
-  (A_{t,j} - \tilde{p}_t (1 \given S_t) ) f_t (S_t)^\top \beta \right)^2 \right]
\]
Differentiating the above equation with respect to~$\alpha$
yields a set of~$p$ estimating equations.
\begin{align*}
0_{q^\prime}&= \E \left[ \frac{1}{G} \sum_{j=1}^G \sum_{t=1}^T W_{t,j} \times W_{t,\Delta,j} \left( Y_{t,\Delta,j} - g_t(H_t)^\top \alpha -  (A_{t,j} - \tilde{p}_t (1 \given S_t) ) f_t (S_t)^\top \beta \right) g_t(H_t) \right]
\end{align*}
We note that
$$
\E \left[ W_{t,J} W_{t,\Delta, J} (A_{t,J} - \tilde{p}_t (1 \given S_t) )
f_t (S_t)^\top \beta \given H_t \right] = 0.
$$
Therefore, we have,
\begin{align*}
0_{p}&= \E \left[ \frac{1}{G} \sum_{j=1}^G \sum_{t=1}^T (g_t(H_t) \E \left[ W_{t,j} W_{t,\Delta, j} Y_{t,\Delta, j} \mid H_t \right] - g_t(H_t)  g_t(H_t)^\top \alpha) \right] \\
\Rightarrow \alpha  &= \E \left[ \frac{1}{G} \sum_{j=1}^G \sum_{t=1}^T g_t(H_t)  g_t(H_t)^\top  \right]^{-1} E \left[ \frac{1}{G} \sum_{j=1}^G \sum_{t=1}^T g_t(H_t) \E \left[ W_{t,j} W_{t,\Delta, j} Y_{t,\Delta, j} \mid H_t \right] \right]
\end{align*}
We note that
\begin{align*}
&\E \left[ W_{t,J} W_{t,\Delta, J} (A_{t,J} - \tilde{p}_t (1 \given S_t) )
g_t (H_t)^\top \alpha \right] = 0, \quad \text{and} \\
&\E \left[ W_{t,J} W_{t,\Delta, J} (A_{t,J} - \tilde{p}_t (1 \given S_t) )
Y_{t,\Delta,J} \right] =
\tilde{p}_t (1 \given S_t) (1- \tilde{p}_t (1 \given S_t)) \beta_{{\bf p}, \pi, \Delta} (t; S_t) , \quad \text{and} \\
&\E \left[ W_{t,J} W_{t,\Delta, J} (A_{t,J} - \tilde{p}_t (1 \given S_t) )^2 \mid S_t
\right] =
\tilde{p}_t (1 \given S_t) (1- \tilde{p}_t (1 \given S_t))
\end{align*}
Now differentiating with respect to~$\beta$ yields
\begin{align*}
0_{q}&= \E \left[ \sum_{t=1}^T W_{t,J} \left( Y_{t,\Delta,J} - g_t(H_t)^\top \alpha
-  (A_{t,J} - \tilde{p}_t (1 \given S_t) ) f_t (S_t)^\top \beta \right) (A_{t,J} - \tilde{p}_t(1 \given S_t)) f_t (S_t) \right] \\
0_{q}&= \E \left[ \sum_{t=1}^T \tilde{p}_t (1 \given S_t) (1- \tilde{p}_t (1 \given S_t)) \left( \beta_{{\bf p}, \pi, \Delta} (t; S_t) - f_t (S_t)^\top \beta^\star \right) f_t (S_t) \right]
\end{align*}
Then we have
\begin{align*}
\beta^\star &= \E\left[ \sum_{t=1}^T  \tilde{p}_t (1 \given S_t) (1- \tilde{p}_t (1 \given S_t)) f_t(S_t) f_t(S_t)^\top \right]^{-1} \E\left[ \sum_{t=1}^T  \tilde{p}_t (1 \given S_t) (1- \tilde{p}_t (1 \given S_t)) f_t(S_t) \beta_{{\bf p}, \pi, \Delta} (t;S_t) \right]
\end{align*}
Under assumption~\ref{ass:directeffect}, we have that $\beta = \beta^\star$ which guarantees consistency.

We next consider the indirect effect estimator.  Recall that
$$
\tilde{p}^\star_t (1 \given S_t) = \frac{\tilde{p}_t (0,1 \given S_t)}{\tilde{p}_t (0,0 \given S_t)+\tilde{p}_t (0,1 \given S_t)}
$$
is the replacement for $\tilde p(1 \mid S_t)$ in the direct effect for centering.  If we make the assumption that $\tilde p_t (0,1 \mid S_t) = \tilde p_t (0 \mid S_t) \tilde p_t (1 \mid S_t)$ then $\tilde{p}^\star_t (1 \given S_t) = \tilde{p}_t (1 \given S_t)$; however, we provide the proof in complete generality. The estimates that minimize equation~\eqref{eq:indirectwcls} are consistent estimators for the solutions that minimize the following
\[
\E \left[ \sum_{t=1}^T W_{t,J, J^\prime} W_{t,\Delta, J,J^\prime} \left( Y_{t,\Delta,J} - g_t(H_t)^\top \alpha
-  (1-A_{t,J}) (A_{t,J^\prime} - \tilde{p}^\star_t (1 \given S_t) ) f_t (S_t)^\top \beta \right)^2 \right]
\]
Differentiating the above equation with respect to~$\alpha$
yields a set of~$p$ estimating equations.
\begin{align*}
0_{p}&= \E \left[ \sum_{t=1}^T W_{t,J,J^\prime} W_{t,\Delta, J,J^\prime} \left( Y_{t,\Delta,J} - g_t(H_t)^\top \alpha
-  (1-A_{t,J}) (A_{t,J^\prime} - \tilde{p}_t^\star (1 \given S_t) ) f_t (S_t)^\top \beta \right) g_t(H_t) \right]
\end{align*}
We note that
$$
\E \left[ W_{t,J, J^\prime} W_{t,\Delta, J,J^\prime} (1-A_{t,J}) (A_{t,J^\prime} - \tilde{p}_t^\star (1 \given S_t) )
f_t (S_t)^\top \beta \given H_t \right] = 0.
$$
Therefore, we have,
\begin{align*}
0_{p}&= \E \left[ \sum_{t=1}^T (g_t(H_t) \E \left[ W_{t,J,J^\prime} W_{t,\Delta, J,J^\prime} Y_{t,\Delta, J} \mid H_t \right] - g_t(H_t)  g_t(H_t)^\top \alpha) \right] \\
\Rightarrow \alpha  &= \E \left[ \frac{1}{G} \sum_{j=1}^G \sum_{t=1}^T g_t(H_t)  g_t(H_t)^\top  \right]^{-1} E \left[ \sum_{t=1}^T g_t(H_t) \E \left[ W_{t,J, J^\prime} W_{t,\Delta, J,J^\prime} Y_{t,\Delta, J} \mid H_t \right] \right]
\end{align*}
First, we show that
\begin{align*}
&\E \left[ W_{t,J, J^\prime} W_{t,\Delta, J,J^\prime} (1-A_{t,J}) (A_{t,J^\prime} - \tilde{p}_t^\star (1 \given S_t) )
\mid H_t \right] \\
= & \E \left[ W_{t,J, J^\prime} (1-A_{t,J}) (A_{t,J^\prime} - \tilde{p}_t^\star (1 \given S_t) )
\mid H_t \right] \\
= &\sum_{a^\prime \in \{0,1\}} \E \left[ \tilde{p}_t (0,a^\prime \mid S_t)  (a^\prime - \tilde{p}_t^\star (1 \given S_t) )
\mid H_t, A_t = 0, A_{t,J^\prime} = a^\prime \right]\\
=& \tilde{p}_t (0,1 \mid S_t)  (1 - \frac{\tilde{p}_t (0,1 \given S_t)}{\tilde{p}_t (0,0 \given S_t)+\tilde{p}_t (0,1 \given S_t)} ) -  \tilde{p}_t (0,0 \mid S_t) \frac{\tilde{p}_t (0,1 \given S_t)}{\tilde{p}_t (0,0 \given S_t)+\tilde{p}_t (0,1 \given S_t)} = 0
\end{align*}
and
\begin{align*}
&\E \left[ W_{t,J, J^\prime} W_{t,\Delta, J,J^\prime} (1-A_{t,J}) (A_{t,J^\prime} - \tilde{p}_t^\star (1 \given S_t) )^2
\mid H_t \right] \\
= &\E \left[ W_{t,J, J^\prime} (1-A_{t,J}) (A_{t,J^\prime} - \tilde{p}_t^\star (1 \given S_t) )^2
\mid H_t \right] \\
=& \tilde{p}_t (0,1 \mid S_t) \left( \frac{\tilde{p}_t (0,0 \given S_t)}{\tilde{p}_t (0,0 \given S_t) +\tilde{p}_t (0,1 \given S_t)} \right)^2 + \tilde{p}_t (0,0 \mid S_t) \left( \frac{\tilde{p}_t (0,1 \given S_t)}{\tilde{p}_t (0,0 \given S_t)+\tilde{p}_t (0,1 \given S_t)} \right)^2 \\
=& ( \tilde{p}_t (0,0 \mid S_t) + \tilde{p}_t (0,1 \mid S_t) ) \tilde{p}_t^\star (1 \mid S_t) (1- \tilde{p}_t^\star (1 \mid S_t)).
\end{align*}
This implies that
\begin{align*}
&\E \left[ W_{t,J, J^\prime} W_{t,\Delta, J,J^\prime} (1-A_{t,J}) (A_{t,J^\prime} - \tilde{p}_t^\star (1 \given S_t) )
g_t (H_t)^\top \alpha \right] = 0, \quad \text{and} \\
&\frac{\E \left[ W_{t,J, J^\prime}  W_{t,\Delta, J,J^\prime} (1-A_{t,J}) (A_{t,J^\prime} - \tilde{p}_t^\star (1 \given S_t) )
Y_{t,\Delta,J} \right]}{\tilde{p}_t (0,0 \mid S_t) + \tilde{p}_t (0,1 \mid S_t)} = \tilde{p}_t^\star (1 \mid S_t) (1- \tilde{p}_t^\star (1 \mid S_t)) \beta^{(IE)}_{{\bf p}, \pi, \Delta} (t; S_t).
\end{align*}
Now differentiating with respect to~$\beta$ yields
\begin{align*}
0_{q}&= \E \left[ \sum_{t=1}^T ( \tilde{p}_t (0,0 \mid S_t) + \tilde{p}_t (0,1 \mid S_t) ) \tilde{p}_t^\star (1 \mid S_t) (1- \tilde{p}_t^\star (1 \mid S_t)) \left( \beta^{(IE)}_{{\bf p}, \pi, \Delta} (t; S_t) - f_t (S_t)^\top \beta^{\star \star} \right) f_t (S_t) \right]
\end{align*}
Under assumption~\ref{ass:indirecteffect}, we have that $\beta = \beta^{\star \star}$ which guarantees consistency.
\end{proof}

\begin{proof}[Asymptotic Normality]
We now consider the issue of asymptotic normality.  First, let
\[
\epsilon_{t,\Delta,j} = Y_{t,\Delta,j} - g_t(H_t)^\top \alpha^\star - (A_{t,j}-\tilde{p}_t(1 \mid S_t)) f_t(S_t)^{\top} \beta^\star,
\]
$\hat{\theta} = (\hat{\alpha}, \hat{\beta})$, and $\theta^\star = (\alpha^\star, \beta^\star)$.
Since $S_t \subset H_t$ define $h_{t,j}(H_t)^\top = (g_t(H_t)^\top, (A_{t,j}-\tilde{p}_t(1 \given S_t)) f_t(S_t)^\top)$. Then
\begin{align*}
\sqrt{M} ( \hat{\theta} - \theta^\star )
  &= \sqrt{M} \bigg \{ \mathbb{P}_M \bigg( \frac{1}{G_m} \sum_{j=1}^{G_m} \sum_{t=1}^T W_{t,j} W_{t,\Delta, j} h_{t,j}(H_t) h_{t,j}(H_t)^\top \bigg)^{-1} \\
  \bigg[
    &\mathbb{P}_M \bigg( \frac{1}{G_m} \sum_{j=1}^{G_m} \sum_{t=1}^T W_{t,j}
    W_{t,\Delta,j} Y_{t,\Delta,j} h_{t,j}(H_t) \bigg) \\
  &- \mathbb{P}_M \bigg( \frac{1}{G_m} \sum_{j=1}^{G_m}
  \sum_{t=1}^T W_{t,j} W_{t,\Delta,j} h_{t,j}(H_t) h_{t,j}(H_t)^\top \bigg)
    \theta^\star \bigg] \bigg \} \\
  &= \sqrt{M} \bigg \{ E \bigg[
  \frac{1}{G} \sum_{j=1}^G \sum_{t=1}^T W_{t,j} W_{t,\Delta,j} h_{t,j}(H_t)
  h_{t,j}(H_t)^\top \bigg]^{-1} \\
  &\bigg[ \mathbb{P}_M \bigg( \frac{1}{G_m} \sum_{j=1}^{G_m} \sum_{t=1}^T
    W_{t,j} W_{t,\Delta,j} \epsilon_{t,\Delta, j} h_{t,j}(H_t) \bigg) \bigg] \bigg \} + o_p ( {\bf 1} )
\end{align*}
By definitions of $\alpha^\star$ and $\beta^\star$ and the previous consistency argument
\[
E \bigg[
  \frac{1}{G} \sum_{j=1}^G \sum_{t=1}^T W_{t,j} W_{t,\Delta, j} h_{t,j}(H_t)
  h_{t,j}(H_t)^\top \bigg]  = 0
\]
Then under moments conditions, we have asymptotic normality with variance $\Sigma_{\theta}$ given by
\begin{align*}
\Sigma_{\theta} &= E \left[ \frac{1}{G} \sum_{j=1}^G \sum_{t=1}^T W_{t,j} W_{t,\Delta,j} h_{t,j}(H_t) h_{t,j}(H_t)^\top \right]^{-1} \\
                &E \left[ \frac{1}{G} \sum_{j=1}^G \sum_{t=1}^T W_{t,j} W_{t,\Delta,j} \epsilon_{t,\Delta,j} h_{t,j}(H_t)
                  \times  \frac{1}{G} \sum_{j=1}^G \sum_{t=1}^T W_{t,j} W_{t,\Delta,j}  \epsilon_{t,\Delta, j} h_{t,j}(H_t)^\top \right] \\
                &E \left[ \frac{1}{G} \sum_{j=1}^G \sum_{t=1}^T W_{t,j} W_{t,\Delta,j}  h_{t,j}(H_t) h_{t,j}(H_t)^\top \right]^{-1}
\end{align*}
Due to centering, the expectation of the matrix
$W_{t,J} h_{t,J}(H_t) h_{t,J} (H_t)^\top$ is block diagonal and
the sub-covariance matrix~$\Sigma_{\beta}$ can be extracted and is equal to
\begin{align*}
 \Sigma_{\beta} &=  \left[ \sum_{t=1}^T E[ (A_{t,J} - \tilde{p}_t (1 \mid S_t)
                  )^2 W_{t,J} W_{t,\Delta,J} f_t (S_t) f_t (S_t)^\top ] \right]^{-1} \\
  &\cdot E \bigg[ \sum_{t=1}^T W_{t,J} W_{t,\Delta,J} \epsilon_{t,\Delta, J}
                  (A_{t,J} - \tilde{p}_t( 1 \mid S_t)) f_t(S_t)
          \times  \sum_{t=1}^T W_{t,J}
          W_{t,\Delta,J} \epsilon_{t,\Delta, J}
                  (A_{t,J} - \tilde{p}_t( 1 \mid S_t)) f_t(S_t)^\top
                  \bigg] \\
 &\, \cdot \left[ \sum_{t=1}^T E[ (A_{t,J} - \tilde{p}_t (1 \mid S_t)
                  )^2 W_{t,J} W_{t,\Delta,J} f_t (S_t) f_t (S_t)^\top ] \right]^{-1}
\end{align*}
The outer terms are equal to $\left[ \sum_{t=1}^T E[ \tilde{p}_t (1 \mid S_t) (1-\tilde{p}_t (1 \mid S_t)) f_t (S_t) f_t (S_t)^\top ] \right]$, which gives us the covariance as desired.

We next consider asymptotic normality in the indirect setting.  First, let
\[
\epsilon_{t,\Delta, j,j^\prime} = Y_{t,\Delta,j} - g_t(H_t)^\top \alpha^{\star \star} - (1 - A_{t,j})(A_{t,j^\prime}-\tilde{p}^\star_t(1 \mid S_t)) f_t(S_t)^{\top} \beta^{\star \star},
\]
$\hat{\theta} = (\hat{\alpha}, \hat{\beta})$, and $\theta^\star = (\alpha^{\star \star}, \beta^{\star \star})$.
Since $S_t \subset H_t$ define $h_{t,j,j^\prime}(H_t)^\top = (g_t(H_t)^\top, (1-A_{t,j}) (A_{t,j^\prime}-\tilde{p}^\star_t(1 \given S_t)) f_t(S_t)^\top)$. Then $\sqrt{M} ( \hat{\theta} - \theta^{\star \star} )$ equals
\begin{align*}
  &\sqrt{M} \bigg \{ \mathbb{P}_M \bigg( \frac{1}{G_m \cdot (G_m-1)} \sum_{j=1}^{G_m} \sum_{j^\prime \neq j} \sum_{t=1}^T W_{t,j,j^\prime} W_{t,\Delta, j,j^\prime}  h_{t,j,j^\prime}(H_t) h_{t,j,j^\prime}(H_t)^\top \bigg)^{-1} \\
  &\bigg[
    \mathbb{P}_M \bigg( \frac{1}{G_m \cdot (G_m - 1)} \sum_{j=1}^{G_m} \sum_{j^\prime \neq j} \sum_{t=1}^T W_{t,j,j^\prime}
    W_{t,\Delta,j,j^\prime} Y_{t,\Delta,j} h_{t,j,j^\prime}(H_t) \bigg)  \\
  &- \mathbb{P}_M \bigg( \frac{1}{G_m \cdot (G_m - 1)} \sum_{j=1}^{G_m}\sum_{j^\prime \neq j}
  \sum_{t=1}^T W_{t,j,j^\prime} W_{t,\Delta,j,j^\prime} h_{t,j, j^\prime}(H_t) h_{t,j, j^\prime}(H_t)^\top \bigg)
    \theta^\star \bigg] \bigg \} \\
  = &\sqrt{M} \bigg \{ E \bigg[
  \frac{1}{G (G-1)} \sum_{j=1}^G \sum_{j^\prime \neq j} \sum_{t=1}^T W_{t,j,j^\prime} W_{t,\Delta,j,j^\prime} h_{t,j, j^\prime}(H_t) h_{t,j, j^\prime}(H_t)^\top \bigg]^{-1} \\
  &\bigg[ \mathbb{P}_M \bigg( \frac{1}{G_m \cdot (G_m-1)} \sum_{j=1}^{G_m} \sum_{j^\prime \neq j} \sum_{t=1}^T W_{t,j,j^\prime} W_{t,\Delta,j,j^\prime} \epsilon_{t,\Delta, j, j^\prime} h_{t,j,j^\prime}(H_t) \bigg) \bigg] \bigg \} + o_p ( {\bf 1} )
\end{align*}
By definitions of $\alpha^{\star \star}$ and $\beta^{\star \star}$  and the previous consistency argument
\[
E \bigg[
  \frac{1}{G \cdot (G-1)} \sum_{j=1}^G \sum_{j^\prime \neq j} \sum_{t=1}^T W_{t,j,j^\prime} W_{t,\Delta,j,j^\prime} h_{t,j,j^\prime}(H_t)
  h_{t,j,j^\prime}(H_t)^\top \bigg]  = 0
\]
Then under moments conditions, we have asymptotic normality with variance $\Sigma_{\theta}$ given by
\begin{align*}
\Sigma_{\theta} &= E \left[ \frac{1}{G \cdot (G-1)} \sum_{j=1}^G \sum_{j^\prime \neq j} \sum_{t=1}^T W_{t,j,j^\prime}
W_{t,\Delta,j,j^\prime} h_{t,j,j^\prime}(H_t) h_{t,j,j^\prime}(H_t)^\top \right]^{-1} \\
                &E \bigg[ \frac{1}{G (G-1)} \sum_{j=1}^G \sum_{j^\prime \neq j} \sum_{t=1}^T W_{t,j, j^\prime} W_{t,\Delta,j,j^\prime} \epsilon_{t, \Delta,j, j^\prime} h_{t,j, j^\prime}(H_t) \\
            &\times  \frac{1}{G(G-1)} \sum_{j=1}^G \sum_{j^\prime \neq j} \sum_{t=1}^T W_{t,j} W_{t,\Delta,j,j^\prime}
                  \epsilon_{t,\Delta, j, j^\prime} h_{t,j, j^\prime}(H_t)^\top \bigg] \\
                &E \left[ \frac{1}{G \cdot (G-1)} \sum_{j=1}^G \sum_{j^\prime \neq j} \sum_{t=1}^T W_{t,j,j^\prime} W_{t,\Delta,j,j^\prime} h_{t,j,j^\prime}(H_t)
  h_{t,j,j^\prime}(H_t)^\top  \right]^{-1}
\end{align*}
Due to centering, the expectation of the matrix
$W_{t,J, J^\prime} h_{t,J,J^\prime}(H_t) h_{t,J, J^\prime} (H_t)^\prime$ is block diagonal and
the sub-covariance matrix~$\Sigma_{\beta}$ can be extracted and is equal to
\begin{align*}
 \Sigma_{\beta} &=  \left[ \sum_{t=1}^T E[ (1- A_{t,J}) (A_{t,J^\prime} - \tilde{p}_t^\star (1 \mid S_t)
                  )^2 W_{t,J, J^\prime} W_{t,\Delta, J, J^\prime} f_t (S_t) f_t (S_t)^\top ] \right]^{-1} \\
  &\cdot E \bigg[ \sum_{t=1}^T W_{t,J, J^\prime} W_{t,\Delta, J, J^\prime} \epsilon_{t, \Delta, J, J^\prime}
                  (1-A_{t,J}) (A_{t,J^\prime} - \tilde{p}_t( 1 \mid S_t)) f_t(S_t) \\
          &\times  \sum_{t=1}^T W_{t,\tilde  J, \tilde J^\prime} W_{t,\Delta, \tilde J, \tilde J^\prime} \epsilon_{t, \Delta,  \tilde J, \tilde J^\prime}
                  (1-A_{t,\tilde J}) (A_{t,\tilde J^\prime} - \tilde{p}_t^\star( 1 \mid S_t)) f_t(S_t)^\top
                  \bigg] \\
 &\, \cdot \left[ \sum_{t=1}^T E[ (1-A_{t,J}) (A_{t,J^\prime} - \tilde{p}_t (1 \mid S_t)
                  )^2 W_{t,J,J^\prime} W_{t,\Delta, J, J^\prime} f_t (S_t) f_t (S_t)^\top ] \right]^{-1}
\end{align*}
as desired.
\end{proof}

\newpage

\section{Additional simulation details}
\label{app:simdetails}

\begin{table}[!th]
\def~{\hphantom{0}}
\caption{Additional simulation results: cluster-based weighted-centered least squares (C-WCLS) and weighted-least squares estimator (WCLS) comparison for Scenarios I, II, III, and IV.}
\begin{tabular}{lccccccc}
\\
% \hline
Scenario & Estimator & \# of Clusters & Cluster Size & Estimate & SE & RMSE & CP \\ %\hline
\multirow{12}{*}{I}
& C-WCLS & \multirow{2}{*}{25} & \multirow{2}{*}{10} & -0.198 & 0.035 & 0.036  & 0.945 \\
& WCLS & & &  -0.200 & 0.036 & 0.035 & 0.956 \\  %\cdashline{2-8}
& C-WCLS & \multirow{2}{*}{25} & \multirow{2}{*}{25} & -0.199 & 0.022 & 0.023 & 0.948 \\
& WCLS & & &  -0.199 & 0.023 & 0.022 & 0.958 \\ %\cdashline{2-8}
& C-WCLS & \multirow{2}{*}{50} & \multirow{2}{*}{10} & -0.198 & 0.025 & 0.027 & 0.935 \\
& WCLS & & &  -0.198 & 0.026 & 0.026 & 0.944 \\ %\cdashline{2-8}
& C-WCLS & \multirow{2}{*}{50} & \multirow{2}{*}{25} & -0.198 & 0.016 & 0.016 & 0.950 \\
& WCLS & & &  -0.198 & 0.016 & 0.017 & 0.937 \\ %\cdashline{2-8}
& C-WCLS & \multirow{2}{*}{100} & \multirow{2}{*}{10} & -0.199 & 0.018 & 0.018 & 0.949 \\
& WCLS & & &    -0.198& 0.018 & 0.019 & 0.949 \\ %\cdashline{2-8}
& C-WCLS & \multirow{2}{*}{100} & \multirow{2}{*}{25} &  -0.198 & 0.011 & 0.012  & 0.941 \\
& WCLS & & &  -0.199 & 0.011 &  0.012 & 0.944 \\ %\hline
& &&& & & & \\
\multirow{12}{*}{II}
& C-WCLS & \multirow{2}{*}{25} & \multirow{2}{*}{10} & -0.199 & 0.070 & 0.076 & 0.935 \\
& WCLS & & &  -0.201 & 0.041 & 0.076 & 0.710 \\  %\cdashline{2-8}
& C-WCLS & \multirow{2}{*}{25} & \multirow{2}{*}{25} & -0.196 & 0.065 & 0.071 &   0.933 \\
& WCLS & & &  -0.200 & 0.026 & 0.065 & 0.557 \\ %\cdashline{2-8}
& C-WCLS & \multirow{2}{*}{50} & \multirow{2}{*}{10} & -0.200 & 0.051 & 0.049 & 0.957 \\
& WCLS & & &  -0.200 & 0.029 & 0.052 & 0.723 \\ %\cdashline{2-8}
& C-WCLS & \multirow{2}{*}{50} & \multirow{2}{*}{25} & -0.200 & 0.047 & 0.049 & 0.947 \\
& WCLS & & &  -0.199 & 0.019 & 0.048 & 0.555 \\ %\cdashline{2-8}
& C-WCLS & \multirow{2}{*}{100} & \multirow{2}{*}{10} & -0.198  & 0.036 & 0.035 & 0.955 \\
& WCLS & & &  -0.198 & 0.021 & 0.037 & 0.718 \\ %\cdashline{2-8}
& C-WCLS & \multirow{2}{*}{100} & \multirow{2}{*}{25} & -0.198  & 0.033 & 0.035 & 0.942 \\
& WCLS & & &  -0.199 & 0.013 & 0.032 & 0.583 \\ %\hline
& &&& & & & \\
\multirow{12}{*}{III}
& C-WCLS & \multirow{2}{*}{25} & \multirow{2}{*}{10} & -0.199 & 0.070 & 0.073 & 0.948 \\
& WCLS & & &  -0.202 & 0.041 & 0.074 & 0.734 \\  %\cdashline{2-8}
& C-WCLS & \multirow{2}{*}{25} & \multirow{2}{*}{25} & -0.196 & 0.066 & 0.069 & 0.931 \\
& WCLS & & &  -0.200 & 0.026 & 0.068 & 0.563 \\ %\cdashline{2-8}
& C-WCLS & \multirow{2}{*}{50} & \multirow{2}{*}{10} & -0.198 & 0.051 & 0.052 & 0.941 \\
& WCLS & & &  -0.199 &  0.029 & 0.052 & 0.742 \\ %\cdashline{2-8}
& C-WCLS & \multirow{2}{*}{50} & \multirow{2}{*}{25} & -0.199 & 0.047 & 0.048 & 0.946 \\
& WCLS & & &  -0.200 & 0.018 & 0.048 & 0.561 \\ %\cdashline{2-8}
& C-WCLS & \multirow{2}{*}{100} & \multirow{2}{*}{10} & -0.200 & 0.036 & 0.037 & 0.946 \\
& WCLS & & &    -0.200 & 0.021 & 0.037 & 0.740 \\ %\cdashline{2-8}
& C-WCLS & \multirow{2}{*}{100} & \multirow{2}{*}{25} & -0.201 & 0.034 & 0.033 & 0.956 \\
& WCLS & & &  -0.198 & 0.013 & 0.034 & 0.555 \\ %\hline
\end{tabular}
\label{tab:simresults_appendix}
%\begin{tabnote}
%U.S., United States of America; R, respondent.
%\end{tabnote}
\end{table}

\begin{table}[!th]
\def~{\hphantom{0}}
\caption{Simulations show strong finite sample estimation and accurate coverage for indirect effects.}
\begin{tabular}{lcccccc}
\\
% \hline
Scenario & \# of Clusters & Cluster Size & Estimate & SD & RMSE & CP \\ %\hline
\multirow{6}{*}{IV} & 25 & 10 & -0.097 & 0.028 & 0.029 & 0.958 \\
& 25 & 25 & -0.101 & 0.017 &  0.019 & 0.942 \\
& 50 & 10 & -0.097 & 0.020 & 0.021 & 0.953 \\
& 50 & 25 & -0.100 & 0.013 & 0.013 & 0.942 \\
& 100 & 10 & -0.097 & 0.015 & 0.015 & 0.943 \\
& 100 & 25 &  -0.100 & 0.009 & 0.009 & 0.944 \\ %\hline
\end{tabular}
\label{tab:simresults_indirect}
\end{table}

\section{Simulation for Lag Effect estimation}
\label{app:lagsimulation}
\subsection{Simulation setup}

To evaluate the proposed estimator with $\Delta > 1$, we extend the simulation setup in the main paper. Consider an MRT with the same setting, in addition to $\beta_{\Delta 0} =-0.1$ and $\beta_{\Delta 1} = 0.2$, thus, $\Delta = 2$ and the proximal response is:
\begin{equation}
\label{eq:laggenerativemodel}
\begin{array}{r@{}l}
    Y_{t,2} =  \theta_1 \{S_{t+1}&{} - \E \left[ S_{t+1}|A_{t},H_{t}\right]\} + \{A_{t}-p_{t}(1|H_{t})\} (\beta_{\Delta 0}+\beta_{\Delta 1} S_{t}) \\
    &{}+ \{A_{t+1} - p_{t+1}(1|H_{t+1})\}(\beta_{10} + \beta_{11} S_{t+1})+  e_{t+2}.
\end{array}
\end{equation}
% \begin{equation}
% \label{eq:laggenerativemodel}
% \begin{array}{r@{}l}
% Y_{t,1,J} = (-0.2 + 0.2 \cdot S_{t,j})&{} \times (A_{t,j} -p_t(1|H_{t,j})) + (-0.1 + 0.2 \cdot S_{t-1,j})  \times \\ (A_{t-1,j} &{}-p_{t-1}(1|H_{t-1,j})) + 0.8 S_{t,j} + e_g +e_{t+1,j}
% \end{array}
% \end{equation}

Here we identify two prespecified future (after time t) "reference" treatment regimes that define the distribution for $A_{t+1},\dots,A_{t+\Delta-1}$. The first one assigns treatment with probabilities between zero and one and corresponds to the treatment assignment distribution, and the second one chooses the reference regime $A_u = 1$ for $u>t$, with probability one. In this case, the lag $\Delta$ treatment effect represents the impact of a sequential treatments on the proximal response $\Delta$ time units later.

\noindent {\bf Simulation Scenario I}. The first scenario estimates $\beta_\Delta^*$ when an individual-level moderator exists and proximal responses share a random cluster-level intercept term that does not interact with treatment. The data generative model~\eqref{eq:laggenerativemodel} incorporates a cluster-level random-intercept $e_g \sim \mathcal{N}(0,0.5)$.
% , so that $
% Y_{t,1,J} = (-0.2 + 0.2 \cdot S_{t,j}) \times (A_{t,j} -p_t(1|H_{t,j})) + 0.8 S_{t,j} + e_g +e_{t+1,j}$.
Table~\ref{tab:lagsimresults} presents the results, which shows under both future treatment policies, the proposed C-WCLS approach achieve nearly unbiasedness and proper coverage.
% This demonstrates that the performance of the WCLS approach is not impacted by group-level correlation that does not interact with treatment.

\noindent {\bf Simulation Scenario II}. In the second scenario, we extend scenario I to include a random cluster-level intercept term that interacts with the treatment by considering the linear model with the additional term~$b_g^\prime \times (A_{t,j} -p_{t}(1|H_{t,j}))$
%\begin{equation}
%Y_{t,1,J} = (-0.2 + b_g +  0.2 \cdot S_{t,j}) \times (A_{t,j} -p_t(1|H_{t,j})) + 0.8 \cdot S_{t,j} + e_g + e_{t+1,j}
%\end{equation}
where $b_g^\prime \sim \mathcal{N}(0,0.1)$.
% is a random-intercept term within the treatment effect per cluster.
Table~\ref{tab:lagsimresults} presents the results which demonstrate that if cluster-level random effects interact with the previous treatment, then both policies produce nearly unbiased estimates and the proposed method achieves the nominal 95\% coverage probability.
% This shows that the coverage probability of the WCLS method decays rapidly while the proposed method achieves the nominal 95\% coverage probability for all choices of the variance of $b_g$ and when the group size increases.
% In Figure~\ref{fig:undercoverage} (left panel), note that even when group size is $5$ (i.e., small groups), the nominal coverage drops to 80\%.

\noindent {\bf Simulation Scenario III}.  In the third scenario, the lag treatment effect for an individual is assumed to depend on the average state of all individuals in the cluster, i.e., define the cluster-level moderator $\bar S_{t,g} = \frac{1}{G_g}\sum_{j=1}^{G_g} S_{t,j}$ and consider the linear model from Scenario II with the additional term~$\bar S_{t,g} \times (A_{t,j} -p_{t}(1|H_{t,j}))$.
%\begin{equation*}
% Y_{t,1,J} = (-0.2 + b_g +  0.2 \cdot \bar S_{t,g}) \times (A_{t,j} -p_t(1|H_{t,j})) + 0.8 S_{t,j} + e_g + e_{t+1,j}
% \end{equation*}
The proposed estimator again achieves the nominal 95\% coverage probability (see Scenario III, Table~\ref{tab:lagsimresults}).

\subsection{Lag Treatment Effect Calculation}

\noindent \textbf{Sequential Treatment Regime.}~As stated by the sequential treatment reference regime, we have the weight $W_{t,\Delta} = \frac{\pi(A_{t+1}\mid H_{t+1})}{p(A_{t+1}\mid H_{t+1})} = \frac{1[A_{t+1} = 1]}{p (A_{t+1} | H_{t+1} ) }  $. Thus, the true lag $\Delta=2$ treatment effect can simply be calculated as:
\begin{equation}
    \beta_{t,2}= E \left[ Y_{t,2} \frac{1[A_{t+1} = 1]}{p (A_{t+1} | H_{t+1} ) } \mid H_{t}, A_{t} = 1 \right]  -  E \left[ Y_{t,2} \frac{1[A_{t+1} = 1]}{p (A_{t+1} | H_{t+1} ) } \mid H_{t}, A_{t} = 0 \right]
\end{equation}

Under our simulation setting, the term $E \left[ Y_{t,2} \frac{1[A_{t+1} = 1]}{p (A_{t+1} | H_{t+1} ) } \mid H_{t}, A_{t} = a \right]$ is equal to:
\begin{align*}
    E &\left[ \left( 0.8 S_{t+1} + (A_t - p_t(1| H_t)) (-0.1 + 0.2 S_{t}) + (A_{t+1} - p_{t+1}(1 | H_{t+1})) (-0.2 + 0.2 S_{t+1} ) \right)\right. \\
    &\times \left.\frac{1[A_{t+1} = 1]}{p (A_{t+1} | H_{t+1} ) } \mid H_{t}, A_{t} = a \right]
\end{align*}
Splitting the expectation above to three terms, we have the following calculation:
\begin{equation}
\label{term1}
    E \bigg[ 0.8 S_{t+1} \frac{1[A_{t+1} = 1]}{p (A_{t+1} | H_{t+1} ) } \mid H_{t}, A_{t} = a \bigg] = 0
\end{equation}
\begin{align}
\label{term2}
        &  E \left[(A_t - p_t(1| H_t)) (-0.1 + 0.2 S_{t})\frac{1[A_{t+1} = 1]}{p (A_{t+1} | H_{t+1} ) } \mid H_{t}, A_{t} = a \right] \nonumber \\
        &= (a - p_t(1| H_t)) (-0.1 + 0.2 S_{t}) E \bigg[\frac{1[A_{t+1} = 1]}{p (A_{t+1} | H_{t+1} ) } \mid H_{t}, A_{t} = a \bigg] \nonumber \\
        & = (a - p_t(1| H_t)) (-0.1 + 0.2 S_{t})
\end{align}

\begin{align}
\label{term3}
    & E \left[ (A_{t+1} - p_{t+1}(1 | H_{t+1})) (-0.2 + 0.2 S_{t+1} ) \frac{1[A_{t+1} = 1]}{p (A_{t+1} | H_{t+1} ) } \mid H_{t}, A_{t} = a \right] \nonumber \\
    & = E \left[ (-0.2 + 0.2 S_{t+1} ) E \left[ (A_{t+1} - p_{t+1}(1 | H_{t+1})) \frac{1[A_{t+1} = 1]}{p (A_{t+1} | H_{t+1} ) } \mid H_{t+1} \right] \mid H_{t}, A_{t} = a \right] \nonumber \\
    & = E \left[ (-0.2 + 0.2 S_{t+1}) (1- p_{t+1} (1 |H_{t+1}) ) \mid H_t, A_t = a \right] \nonumber \\
    &=-0.2 + 0.2 \cdot E[(1- S_{t+1}) \cdot  \text{expit}(-0.8 A_{t}+0.8 S_{t+1}) \mid H_t, A_t = a] \nonumber \\
    & = -0.2 + 0.2 \cdot \text{expit}(-0.8a-0.8)
\end{align}

Therefore,
\begin{align*}
    E \left[ Y_{t,2} \frac{1[A_{t+1} = 1]}{p (A_{t+1} | H_{t+1} ) } \mid H_{t}, A_{t} = a \right] & = \text{(\ref{term1})} + \text{(\ref{term2})} + \text{(\ref{term3})} \\
    & = (a - p_t(1| H_t)) (-0.1 + 0.2 S_{t})-0.2 + 0.2 \cdot \text{expit}(-0.8a-0.8)
\end{align*}
and the true lag $\Delta=2$ treatment effect under sequential treatment regime is equal to:
\begin{align}
    \beta_{t,2}&= E \left[ Y_{t,2} \frac{1[A_{t+1} = 1]}{p (A_{t+1} | H_{t+1} ) } \mid H_{t}, A_{t} = 1 \right]  -  E \left[ Y_{t,2} \frac{1[A_{t+1} = 1]}{p (A_{t+1} | H_{t+1} ) } \mid H_{t}, A_{t} = 0 \right] \nonumber \\
    & =-0.1 + 0.2S_t + 0.2 \cdot \text{expit}(-0.8-0.8)- 0.2 \cdot \text{expit}(-0.8) \nonumber \\
    & = -0.128 +0.2S_t
\end{align}

\noindent \textbf{Observed Distribution Treatment Regime.}~As specified by this reference treatment regime, we have future treatment reference distribution the same with the distribution of treatments in the data we have at hand, i.e., $\pi(A_{t+1}\mid H_{t+1}) = p(A_{t+1}\mid H_{t+1})$ and $W_{t,\Delta} =1$. Thus, the true lag $\Delta=2$ treatment effect can be calculated as:
\begin{equation}
    \beta^\prime_{t,2} = E \left[ Y_{t,2} \mid H_{t}, A_{t} = 1 \right] - E \left[ Y_{t,2} \mid H_{t}, A_{t} = 0 \right]
\end{equation}
Similar as above, under our simulation setting, the term $E \left[ Y_{t,2} \mid H_{t}, A_{t} = a \right]$ is equal to:
\begin{align*}
    E &\left[ \left( 0.8 S_{t+1} + (A_t - p_t(1| H_t)) (-0.1 + 0.2 S_{t}) + (A_{t+1} - p_{t+1}(1 | H_{t+1})) (-0.2 + 0.2 S_{t+1} ) \right)   \mid H_{t}, A_{t} = a \right] \\
    &= \text{(\ref{termb1})} + \text{(\ref{termb2})} + \text{(\ref{termb3})}
\end{align*}
The three terms are calculated below:
\begin{equation}
\label{termb1}
    E \left[ 0.8 S_{t+1}   \mid H_{t}, A_{t} = a \right] = 0
\end{equation}
\begin{align}
\label{termb2}
         E \left[(A_t - p_t(1| H_t)) (-0.1 + 0.2 S_{t})  \mid H_{t}, A_{t} = a \right] = (a - p_t(1| H_t)) (-0.1 + 0.2 S_{t})
\end{align}

\begin{align}
\label{termb3}
    & E \left[ (A_{t+1} - p_{t+1}(1 | H_{t+1})) (-0.2 + 0.2 S_{t+1} )   \mid H_{t}, A_{t} = a \right] \nonumber \\
     = &E \left[ (-0.2 + 0.2 S_{t+1} ) E \left[ (A_{t+1} - p_{t+1}(1 | H_{t+1}))   \mid H_{t+1} \right] \mid H_{t}, A_{t} = a \right] \nonumber \\
     = &0
\end{align}

Therefore,
\begin{equation*}
    E \left[ Y_{t,2}   \mid H_{t}, A_{t} = a \right] =  (a - p_t(1| H_t)) (-0.1 + 0.2 S_{t})
\end{equation*}
and the true lag $\Delta=2$ treatment effect under observed treatment distribution regime is equal to:
\begin{align}
    \beta^\prime_{t,2}&= E \left[ Y_{t,2}   \mid H_{t}, A_{t} = 1 \right]  -  E \left[ Y_{t,2}   \mid H_{t}, A_{t} = 0 \right] \nonumber \\
    & =-0.1 + 0.2S_t
\end{align}

\subsection{Marginal Lag Treatment Effect Simulation Results}

The choice for prespecified future reference treatment regimes is of vital importance and often time yields to different treatment effect estimations. Following the derivations above, the fully marginal lag $\Delta=2$ treatment effect is -0.128 for sequential treatment reference regime, and -0.1 for observed treatment distribution regime. Table~\ref{tab:lagsimresults} presents the simulation results.

\begin{table}[!th]
\def~{\hphantom{0}}
\caption{Simulation: cluster-based weighted-centered least squares (C-WCLS) estimators for lag $\Delta =2$ effect, under the policy of observed treatment distribution (OTD) versus sequential treatment (ST), and comparison for Scenarios I, II, III.}
\begin{tabular}{lccccccc}
\\
% \hline
Scenario & Policy & \# of Clusters & Cluster Size & Estimate & SE & RMSE & CP \\ %\hline
\multirow{4}{*}{I}
& OTD & \multirow{2}{*}{50} & \multirow{2}{*}{10} & -0.098 & 0.031 & 0.032 & 0.938 \\
& ST & & &  -0.123 & 0.062 & 0.063 & 0.949 \\ %\cdashline{2-8}
& OTD & \multirow{2}{*}{50} & \multirow{2}{*}{25} & -0.098 & 0.020 & 0.020 & 0.948 \\
& ST & & &  -0.124 & 0.040 & 0.039 & 0.955 \\%\hline %\cdashline{2-8}
& &&& & & & \\
\multirow{4}{*}{II}
& OTD & \multirow{2}{*}{50} & \multirow{2}{*}{10} & -0.099 & 0.054 & 0.054 & 0.944 \\
& ST & & &  -0.122 & 0.077 & 0.078 & 0.949 \\ %\cdashline{2-8}
& OTD & \multirow{2}{*}{50} & \multirow{2}{*}{25} & -0.099 & 0.048 & 0.049 & 0.942 \\
& ST & & &  -0.121 & 0.059 & 0.061 & 0.950 \\ %\hline %\cdashline{2-8}
& &&& & & & \\
\multirow{4}{*}{III}
& OTD & \multirow{2}{*}{50} & \multirow{2}{*}{10} & -0.096 & 0.054 & 0.056 & 0.942 \\
& ST & & &  -0.122 &  0.075 & 0.075 & 0.948 \\ %\cdashline{2-8}
& OTD & \multirow{2}{*}{50} & \multirow{2}{*}{25} & -0.099 & 0.048 & 0.050 & 0.944 \\
& ST & & &  -0.125 & 0.059 & 0.059 & 0.955 \\ %\hline %\cdashline{2-8}
\end{tabular}
\label{tab:lagsimresults}
%\begin{tabnote}
%U.S., United States of America; R, respondent.
%\end{tabnote}
\end{table}

\section{Proof of Lemma~\ref{lemma:samesies}}
\label{app:samesies}

\begin{proof}
Consider the $W$-matrix for the direct effect asymptotic variance,
\begin{align*}
 &\frac{1}{G^2} \sum_{t, t^\prime} \sum_{j, j^\prime} \mathbb{E} \bigg[
 W_{t,j} W_{t,\Delta, j} \epsilon_{t,\Delta, j} (A_{t,j} - \tilde{p}_t( 1 \mid S_t))
 W_{t^\prime,j^\prime}
 W_{t^\prime,\Delta, j^\prime} \epsilon_{t^\prime,\Delta, j^\prime} (A_{t^\prime,j^\prime} - \tilde{p}_t( 1 \mid S_{t^\prime}))
 f_t(S_t) f_{t^\prime}(S_{t^\prime})^\top
                  \bigg]
% =&\frac{1}{G^2} \sum_{t, t^\prime}\sum_{j, j^\prime} \mathbb{E} \bigg[
%  W_{t,j} \epsilon_{t,j} (A_{t,J} - \tilde{p}_t( 1 \mid S_t))
%  W_{t^\prime,j^\prime} \epsilon_{t^\prime,j^\prime} (A_{t^\prime,j^\prime} - \tilde{p}_t( 1 \mid S_{t^\prime}))
%  f_t(S_t) f_{t^\prime}(S_{t^\prime})^\top
%                   \bigg]
\end{align*}
Consider the cross-terms with $j \neq j^\prime$ and without loss of generality assume $t \geq t^\prime$, then
\begin{align*}
\mathbb{E} &\bigg[ \sum_{a, a^\prime}  \tilde p_t (a \mid S_t) (a - \tilde{p}_t( 1 \mid S_t))
\tilde p_{t^\prime} (a^\prime \mid S_{t^\prime}) (a^\prime - \tilde p_{t^\prime} (1 \mid S_{t^\prime})) \\
&\mathbb{E} \bigg[ \mathbb{E} \bigg[ W_{t,\Delta, j} \epsilon_{t,\Delta, j} W_{t^\prime,\Delta, j^\prime} \epsilon_{t^\prime,\Delta, j^\prime} \mid H_{t,j}, A_{t,j} = a, H_{t^\prime,j^\prime}, A_{t^\prime,j^\prime} = a^\prime \bigg] \mid S_t, S_{t^\prime} \bigg] f_t(S_t) f_{t^\prime}(S_{t^\prime})^\top
                  \bigg].
\end{align*}
Under the assumption of the error cross-term being constant in $a$ and $a^\prime$ we can re-write the above as:
\begin{align*}
&= \mathbb{E} \left[ \sum_{a, a^\prime}  \tilde p_t (a \mid S_t) (a - \tilde{p}_t( 1 \mid S_t))
\tilde p_{t^\prime} (a^\prime \mid S_{t^\prime}) (a^\prime - \tilde p_{t^\prime} (1 \mid S_{t^\prime})) \psi (S_t, S_{t^\prime}) f_t (S_t) f_{t^\prime} (S_{t^\prime})^\top \right] \\
&= \mathbb{E} \bigg[ \psi (S_t, S_{t^\prime}) f_t (S_t) f_{t^\prime} (S_{t^\prime})^\top \underbrace{\left( \sum_{a, a^\prime}  \tilde p_t (a \mid S_t) (a - \tilde{p}_t( 1 \mid S_t))
\tilde p_{t^\prime} (a^\prime \mid S_{t^\prime}) (a^\prime - \tilde p_{t^\prime} (1 \mid S_{t^\prime})) \right)}_{=0} \bigg] \\
&= \mathbb{E} \left[ \psi (S_t, S_{t^\prime}) f_t (S_t) f_{t^\prime} (S_{t^\prime})^\top \cdot 0 \right] = 0.
\end{align*}
Therefore, we have that the $W$-matrix simplifies to
\begin{align*}
 &\mathbb{E} \bigg[ \sum_{t=1}^T W_{t,J} \epsilon_{t,J} (A_{t,J} - \tilde{p}_t( 1 \mid S_t)) f_t(S_t) \times \sum_{t=1}^T W_{t,J} \epsilon_{t,J} (A_{t,J} - \tilde{p}_t( 1 \mid S_t)) f_t(S_t)^\top
                  \bigg] \\
 =&\mathbb{E} \bigg[\frac{1}{G}  \sum_{j=1}^G \bigg[ \sum_{t=1}^T W_{t,J} \epsilon_{t,J} (A_{t,J} - \tilde{p}_t( 1 \mid S_t)) f_t(S_t) \times \sum_{t=1}^T W_{t,J} \epsilon_{t,J} (A_{t,J} - \tilde{p}_t( 1 \mid S_t)) f_t(S_t)^\top
                  \bigg] \bigg] \\
 =&\mathbb{E} \bigg[ \sum_{t=1}^T W_{t} \epsilon_{t} (A_{t} - \tilde{p}_t( 1 \mid S_t)) f_t(S_t) \times \sum_{t=1}^T W_{t} \epsilon_{t} (A_{t} - \tilde{p}_t( 1 \mid S_t)) f_t(S_t)^\top
                  \bigg]
\end{align*}
which is the $W$ matrix as in the standard MRT analysis.
\end{proof}

\section{Small sample size adjustment for covariance estimation}
\label{app:ssa}

The robust sandwich covariance estimator~\cite{Mancl2001} for the entire variance matrix is given by $Q^{-1} \Lambda Q^{-1}$.  The first term,~$Q$, is given by
\[
 \sum_{m=1}^M \frac{1}{G_m}\sum_{j=1}^{G_m} D_{j,m}^\top W_{j,m} D_{j,m}
\]
where $D_{j,m}$ is the model matrix for individual~$j$ in group $g$ associated with
equation~\eqref{eq:directwcls}, and $W_{j,m}$ is a diagonal matrix of individual weights.
The middle term~$\Lambda$ is given by
\[
\sum_{m=1}^M \frac{1}{G_m^2}\sum_{i,j=1}^{G_m} D_{i,m}^\top W_{i,m} (I_{i,m} - H_{i,m})^{-1}
e_{i,m} e_{j,m}^\top (I_{j,m} - H_{j,m})^{-1} W_{j,m} D_{j,m}
\]
where $I_i$ is an identity matrix of correct dimension, $e_i$ is the individual-specific residual
vector and
\[
H_{j,m} = D_{j,m}
\left( \sum_{m=1}^M \frac{1}{G_m}\sum_{j=1}^{G_m} D_{j,m}^\top W_{j,m} D_{j,m} \right)^{-1}
D_{j,m}^\top W_{j,m}
\]
From $Q^{-1} \Lambda Q^{-1}$ we extract $\hat{\Sigma}_{\beta}$.

\section{Additional analysis of IHS}
\label{app:IHSadditionalanalysis}

%\subsection{Additional Information of Figure \ref{fig:wcls_moderation_IHS}}
\subsection{Additional Information for Figure 3}
\label{app:moreonFigure3}

Figure 3 presented a visually comparison between WCLS and C-WCLS in terms of the moderation of average previous week's proximal outcomes on the effect of notifications on average weekly mood scores, log step counts, and log sleep counts respectively in IHS. Here we attach the table for numerical comparison.

\begin{table}[!th]
\def~{\hphantom{0}}
\caption{Moderation Analysis with $\beta(t; S_t) = \beta_0 + \beta_1 \cdot Y_{t,j} $ }
\begin{tabular}{llccccc}
\\
%\hline
Outcome & Setting & Variables & Estimate & Std. Error & t-value & p-value \\ %\hline
\multirow{4}{*}{Mood}
& \multirow{2}{*}{WCLS} & $\beta_0$ & 0.369 & 0.086 & 4.267 & 0.000 \\
& & $\beta_1$ & -0.055 & 0.011 & -4.822 & 0.000 \\ %\cline{2-7}
& \multirow{2}{*}{C-WCLS} & $\beta_0$ & 0.421 & 0.214 & 1.968 & 0.053 \\
& & $\beta_1$ & -0.061 & 0.028 & -2.181 & 0.032 \\ %\cline{2-7}
% & \multirow{3}{*}{C-WCLS} & $\beta_0$ & 0.665  & 0.321 & 2.072 & 0.040 \\
% & & $\beta_1$ & -0.067  &   0.037  &  -1.804  & 0.072 \\
% & & $\beta_2$ & -0.027 & 0.019 &  -1.393 & 0.164\\ %\hline
& & & & & & \\
\multirow{4}{*}{Steps}
& \multirow{2}{*}{WCLS} & $\beta_0$ & 0.729 & 0.295 & 2.472 & 0.015 \\
& & $\beta_1$ & -0.037 & 0.015 & -2.484 & 0.015 \\ %\cline{2-7}
& \multirow{2}{*}{C-WCLS} & $\beta_0$ & 0.997 & 0.734 & 1.357 & 0.176 \\
& & $\beta_1$ & -0.049 & 0.037 & -1.343 & 0.181 \\ %\cline{2-7}
% & \multirow{3}{*}{C-WCLS} & $\beta_0$ & 0.234  &1.039 &  0.225 & 0.823 \\
% & & $\beta_1$ & -0.011  &   0.045 &   -0.254 &    0.800 \\
% & & $\beta_2$ & -0.007  &   0.021 &   -0.354 &  0.725 \\ %\hline
& & & & & & \\
\multirow{4}{*}{Sleep}
& \multirow{2}{*}{WCLS} & $\beta_0$ & 1.325 & 0.350 & 3.782 & 0.000 \\
& & $\beta_1$ & -0.068 & 0.017 & -3.916 & 0.000 \\ %\cline{2-7}
& \multirow{2}{*}{C-WCLS} & $\beta_0$ & 1.543 & 0.767 & 2.012 & 0.046 \\
& & $\beta_1$ & -0.081 & 0.039 & -2.082 & 0.039 \\ %\cline{2-7}
% & \multirow{3}{*}{C-WCLS} & $\beta_0$ & -0.073 & 0.887 &-0.082  & 0.935  \\
% & &$\beta_1$ & 0.002  &   0.043  &   0.049 & 0.961 \\
% & &$\beta_2$ & 0.001 & 0.008  & 0.135 & 0.893 \\ %\hline
\end{tabular}
\label{tab:IHS_Figure3}
\end{table}

\subsection{Lagged Treatment Effect}

We implement an extended investigation on the lag $\Delta=2$ treatment effect of the targeted mobile notifications. The same two models:
$\beta_{2}(t; S_t) = \beta_0 + \beta_1 \cdot Y_{t,j} + \beta_2 \bar Y_{t,-j}$ are applied to moderation analyses.

\begin{figure}
  % \figuresize{.5}
  % \figurebox{15pc}{20pc}{}[all1.eps]
  \includegraphics[width = 0.9\textwidth]{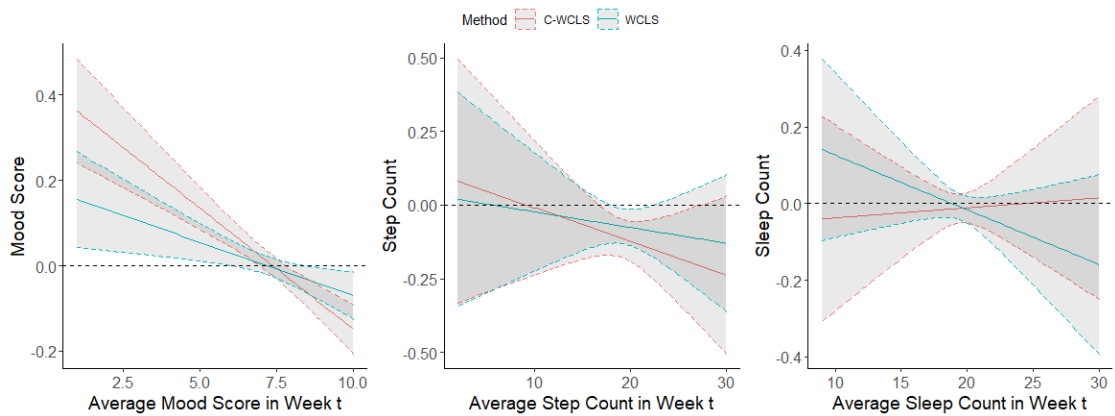}
  \caption{Moderation analysis of lag $\Delta=2$ effect of notifications on average weekly mood scores, log step counts, and log sleep counts respectively under the sequential treatment policy. }
  \label{fig:wcls_heterogeneity1}
\end{figure}

\begin{figure}
\centering
\includegraphics[width=.9\textwidth]{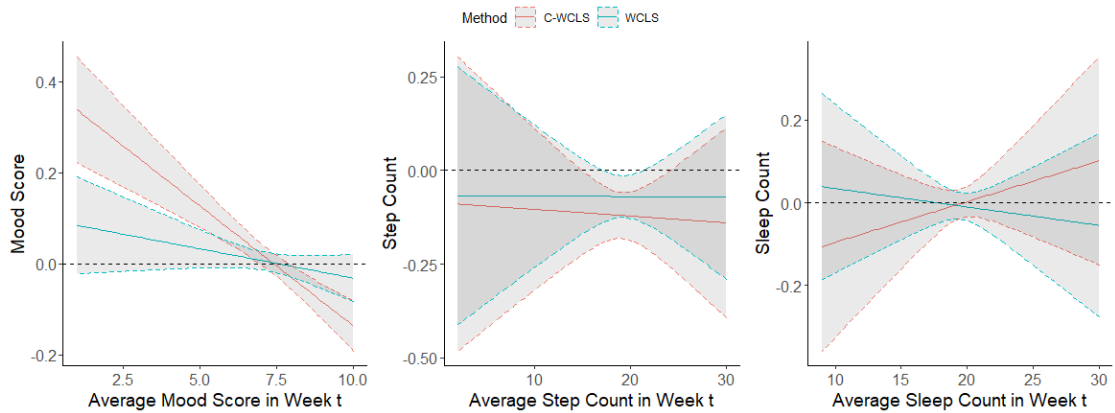}
\caption{Moderation analysis of lag $\Delta=2$ effect of notifications on average weekly mood scores, log step counts, and log sleep counts respectively under the observed treatment distribution. }
\label{fig:wcls_heterogeneity2}
\end{figure}

The first set of moderation analyses considers the standard moderation analysis where only individual-level moderators are included (i.e., $\beta_2 = 0$), with or without accounting for cluster-level moderation effect heterogeneity. Figure~\ref{fig:wcls_heterogeneity1} and Figure~\ref{fig:wcls_heterogeneity2} visualize the estimates across the range of prior week's proximal response for both our proposed approach and the WCLS approach from~\cite{Boruvkaetal}. In this case, the effects do not change too much for all the analysis;

The second moderation analysis lets $\beta_2$ be a free parameter.  Table~\ref{tab:IHS_direct4} and Table~\ref{tab:IHS_direct5} present the results.  Here, we see that the constant term $\beta_0$ in the mood analysis becomes significant, while the new term $\beta_2$ is insignificant.  The results suggest the average proximal outcomes of others in the cluster have a limited moderate effect on notifications, however, the lag impact of a notification on mood is positive and significant.

\begin{table}[!th]
\def~{\hphantom{0}}
\caption{Moderation analysis for lag $\Delta =2$ treatment effect with cluster-level moderators, under the policy of sequential weeks of treatment}
\begin{tabular}{llccccc}
\\
%\hline
Outcome & Setting & Variables & Estimate & Std. Error & t-value & p-value \\ %\hline
\multirow{3}{*}{Mood}
% & \multirow{2}{*}{WCLS} & $\beta_0$ & 0.369 & 0.086 & 4.268 & 0.000 \\
% & & $\beta_1$ & -0.055 & 0.011 & -4.822 & 0.000 \\ \cline{2-7}
% & \multirow{2}{*}{C-WCLS} & $\beta_0$ & 0.350 & 0.103 & 3.401 & 0.001 \\
% & & $\beta_1$ & -0.053 & 0.014 & -3.868 & 0.000 \\ \cline{2-7}
& \multirow{3}{*}{C-WCLS} & $\beta_0$ & 0.665  & 0.321 & 2.072 & 0.040 \\
& & $\beta_1$ & -0.067  &   0.037  &  -1.804  & 0.072 \\
& & $\beta_2$ & -0.027 & 0.019 &  -1.393 & 0.164\\ %\hline
& & & & & & \\
\multirow{3}{*}{Steps}
% & \multirow{2}{*}{WCLS} & $\beta_0$ & 0.729 & 0.295 & 2.472 & 0.015 \\
% & & $\beta_1$ & -0.037 & 0.015 & -2.484 & 0.015 \\ \cline{2-7}
% & \multirow{2}{*}{C-WCLS} & $\beta_0$ & 0.622 & 0.384 & 1.618 & 0.108 \\
% & & $\beta_1$ & -0.031 & 0.019 & -1.580 & 0.117 \\ \cline{2-7}
& \multirow{3}{*}{C-WCLS} & $\beta_0$ & 0.234  &1.039 &  0.225 & 0.823 \\
& & $\beta_1$ & -0.011  &   0.045 &   -0.254 &    0.800 \\
& & $\beta_2$ & -0.007  &   0.021 &   -0.354 &  0.725 \\ %\hline
& & & & & & \\
\multirow{3}{*}{Sleep}
& \multirow{3}{*}{C-WCLS} & $\beta_0$ & -0.073 & 0.887 &-0.082  & 0.935  \\
& &$\beta_1$ & 0.002  &   0.043  &   0.049 & 0.961 \\
& &$\beta_2$ & 0.001 & 0.008  & 0.135 & 0.893 \\ %\hline
\end{tabular}
\label{tab:IHS_direct4}
\end{table}

\begin{table}[!th]
\def~{\hphantom{0}}
\caption{Moderation analysis for lag $\Delta =2$ treatment effect with cluster-level moderators, under the policy of observed treatment distribution}
\begin{tabular}{llccccc}
\\
%\hline
Outcome & Setting & Variables & Estimate & Std. Error & t-value & p-value \\ %\hline
\multirow{3}{*}{Mood}
% & \multirow{2}{*}{WCLS} & $\beta_0$ & 0.369 & 0.086 & 4.268 & 0.000 \\
% & & $\beta_1$ & -0.055 & 0.011 & -4.822 & 0.000 \\ \cline{2-7}
% & \multirow{2}{*}{C-WCLS} & $\beta_0$ & 0.350 & 0.103 & 3.401 & 0.001 \\
% & & $\beta_1$ & -0.053 & 0.014 & -3.868 & 0.000 \\ \cline{2-7}
& \multirow{3}{*}{C-WCLS} & $\beta_0$ & 0.585& 0.292  & 2.002 &  0.046 \\
& & $\beta_1$ & -0.061 & 0.036  &  -1.702 &  0.090 \\
& & $\beta_2$ & -0.022  &   0.021  &  -1.038  & 0.299 \\ %\hline
& & & & & & \\
\multirow{3}{*}{Steps}
% & \multirow{2}{*}{WCLS} & $\beta_0$ & 0.729 & 0.295 & 2.472 & 0.015 \\
% & & $\beta_1$ & -0.037 & 0.015 & -2.484 & 0.015 \\ \cline{2-7}
% & \multirow{2}{*}{C-WCLS} & $\beta_0$ & 0.622 & 0.384 & 1.618 & 0.108 \\
% & & $\beta_1$ & -0.031 & 0.019 & -1.580 & 0.117 \\ \cline{2-7}
& \multirow{3}{*}{C-WCLS} & $\beta_0$ & 0.107 & 0.951  & 0.113  & 0.911 \\
& & $\beta_1$ & -0.001 & 0.041 & -0.033  & 0.974 \\
& & $\beta_2$ & -0.011 &  0.018 &   -0.629  &  0.531  \\ %\hline
& & & & & & \\
\multirow{3}{*}{Sleep}
& \multirow{3}{*}{C-WCLS} & $\beta_0$ & -0.211 & 0.802 & -0.263 &  0.793   \\
& &$\beta_1$ & 0.009 & 0.040 & 0.233 & 0.816 \\
& &$\beta_2$ & 0.001 & 0.007 & 0.193 & 0.848 \\ %\hline
\end{tabular}
\label{tab:IHS_direct5}
\end{table}

\section{Additional details on the indirect effect}
\label{app:addindirect}

Weights used in the estimation of the indirect effect is a natural extension of~\cite{Boruvkaetal}. As in Section~\ref{section:indirect}, the weight~$W_{t,j, j^\prime}$ at decision time $t$ for the $j$th individual is equal to $\frac{\tilde p (A_{t,j}, A_{t,j^\prime} \mid S_t)}{p_t (A_{t,j}, A_{t,j^\prime} \mid H_t)}$ where $\tilde p_t (a, a^\prime \mid S_t)\in (0,1)$ is arbitrary as long as it does not depend on terms in $H_t$ other than $S_t$, and $p(A_{t,j}, A_{t,j^\prime} \mid H_t)$ is the marginal probability that individuals $j$ and $j^\prime$ receive treatments $A_{t,j}$ and $A_{t,j^\prime}$ respectively given $H_t$.

In the simulation, the treatment individuals $j$ and $j^\prime$ receive $A_{t,j}$ and $A_{t,j^\prime}$ are mutually independent conditioning on the previous history. thus, the denominator of $W_{t,j,j\prime}$ can be factorized into:
\[ p(A_{t,j}, A_{t,j^\prime} \mid H_t)= p(A_{t,j} \mid H_t)p(A_{t,j^\prime} \mid H_t) \]
Besides, the numerator of  $W_{t,j,j\prime}$ is defined as the empirical frequency of the treatment pair $ (a, a^\prime)$, which takes the value from $\{(0,0),(0,1),(1,0),(1,1)\}$. Here we denote it as \[\tilde p_t (A_{t,j}, A_{t,j^\prime} \mid S_t) = \hat p_t (A_{t,j}, A_{t,j^\prime} \mid S_t)\]

Therefore, the weight we used in the simulation is constructed as:
\[
W_{t,j,j\prime} = \frac{\hat p_t (A_{t,j}, A_{t,j^\prime} \mid S_t)}{p(A_{t,j} \mid H_t)p(A_{t,j^\prime} \mid H_t)}
\]

When the numerators are estimated using the observed data, the variance-covariance must account for this. Throughout we allow for the setting in which individuals are not always available. For completeness we provide results for a more general estimating function which can be used with observational (non-randomized $A_t$) treatments, under the assumption of sequential ignorability and assuming the data analyst is able to correctly model and estimate the treatment probability $p\left(A_{t,j}, A_{t,j^\prime}| H_t\right)$. We indicate how the results are simplified by use of data from an MRT.

Denote the parameterized treatment probability by $p_t(a,a' | H_t;\eta)$ (with parameter $\eta$); note $\eta$ is known in an MRT. Denote the parameterized numerator of the weights by $\tilde p_t(a,a'| S_t;\rho)$ (with parameter $\rho$). The proof below allows the data analyst to use a  $\tilde p_t$ with an estimated parameter $\tilde \rho$ or to pre-specify $\rho$ as desired. We use a superscript of $\star$ to denote limiting values of estimated parameters (e.g. $\eta^\star, \rho^\star$). Then the more general version of the estimating equation $U_W(\alpha,\beta;\hat\eta,\hat\rho)$ is:
\begin{align*}
    \sum_{t=1}^T \left( Y_{t,1,J} - g_t(H_t)^\top \alpha -  (1-A_{t,J}) (A_{t,J^\prime} - \tilde p_t^\star (1 \given S_t;\hat \rho) ) f_t (S_t)^\top \beta \right)
  &I_{t,J}I_{t,J^\prime}W_{t,J, J^\prime}(A_{t,J},A_{t,J^\prime},H_t; \hat\eta,\hat\rho) \times \\
    &\begin{pmatrix}
  g_t(H_t) \\
  (1-A_{t,J})(A_{t,J^\prime} - \tilde {p}^\star_t (1 \given S_t;\hat\rho) ) f_t (S_t)
\end{pmatrix}
\end{align*}

Note $W_{t,J, J^\prime}$ in the body of the paper is replaced here by $W_{t,J, J^\prime}(A_t,H_t; \hat\eta,\hat\rho)$, and $\hat\eta$, $\hat\rho$ are estimators.

\textbf{Treatment Probability Model:} If the data is observational then we assume: $p_t(a,a'\given{H_t;\eta})$ is a correctly specified model for $\P(a,a'\given{H_t,I_{t,J}=1,I_{t,J^\prime}=1})$. Let $\eta^\star$ be the true value of $\eta$; that is, $Pr(A_{t,J},A_{t,J^\prime}|H_t,I_{t,J}=1,I_{t,J^\prime}=1) = p_t(a,a' \given{H_t;\eta^\star})$ Assume that the estimator of $\eta$, say $\hat\eta$, satisfies $\P_n U_D(\hat\eta)=0$ and $\sqrt{n}(\hat\eta-\eta^\star) =\E \left[\dot U_D (\eta^\star) \right]^{-1}\P_n U_D (\eta^\star) + o_P(1)$. Thus $\sqrt{n}(\hat\eta-\eta^\star)$ converges in distribution to a mean zero, normal random vector with variance-covaraince matrix given by $\E \left[\dot U_D (\eta^\star) \right]^{-1} \E \left[U_D(\eta^\star)^ {\otimes 2} \right]    \left(\E \left[\dot U_D (\eta^\star) \right]^{-1} \right)^\top$, which has finite entries. Assume that $\P_n(\dot U_D(\hat\eta))$ is a consistent estimator of $\E(\dot U_D(\eta^\star))$. Assume there exists finite constants, $b_D>0$ and $B_D<1$ such that each $b_D < p_t (a,a'|H_t;\eta^\star)<B_D$ a.s.

If the data analyst elects to use a parameterized and estimated $\tilde p_t(1|S_t,\hat \rho)$, then we assume:

\textbf{Numerator of Weights Probability Model}: Suppose the estimator $\hat\rho$ solves an estimating equation: $\P_n U_N(\rho)=0$. Assume that, for a finite value of $\rho$, say $\rho^\star$ and $\sqrt{n}(\hat\rho-\rho^\star)= \E \left[\dot U_N (\rho^\star) \right]^{-1}\sqrt{n}(\P_n-P) U_N (\rho^\star) + o_P(1)$ where the matrix, $\E \left[\dot U_N (\rho^\star) \right]$ is positive definite. Assume $\sqrt{n}(\P_n-P) U_N (\rho^\star)$ converges in distribution to a mean zero, normal random vector with variance-covariance matrix given by $\E[U_N(\rho^\star)^ {\otimes 2}]$ which has finite entries. Assume that $\P_n \dot U_N(\hat\rho)$ is a consistent estimator of $\E[\dot U_N(\rho^\star)]$. Assume $0 <\rho^\star <1$.

\begin{proof}
The solution to $\P_n U_W(\alpha,\beta;\hat\eta,\hat\rho) = 0$ gives the estimator:
\begin{align*}
    \begin{pmatrix}
  \hat \alpha \\
  \hat \beta
\end{pmatrix} =
\left\{ \P_n \dot U_W(\hat\eta,\hat\rho) \right \}^{-1} \P_n \sum_{t=1}^T &I_{t,J}I_{t,J^\prime} W_{t,J, J^\prime}(A_{t,J},A_{t,J^\prime},H_t; \hat \eta,\hat \rho) Y_{t,1,J} \times \\
&\begin{pmatrix}
  g_t(H_t) \\
  (1-A_{t,J})(A_{t,J^\prime} - \tilde {p}^\star_t (1 \given S_t) ) f_t (S_t)
\end{pmatrix}
\end{align*}

where

\[
\dot U_W(\hat\eta,\hat\rho) = \sum_{t=1}^T I_{t,J}I_{t,J^\prime}W_{t,J, J^\prime}(A_{t,J},A_{t,J^\prime},H_t; \hat\eta,\hat\rho) \begin{pmatrix}
  g_t(H_t) \\
  (1-A_{t,J})(A_{t,J^\prime} - \tilde {p}^\star_t (1 \given S_t) ) f_t (S_t)
\end{pmatrix}^ {\otimes 2}
\]

Define

\begin{align*}
    \begin{pmatrix}
  \alpha^\prime \\
  \beta^\prime
\end{pmatrix} = \left\{ \E \left[ \dot U_W(\eta^\star,\rho^\star) \right]\right\}^{-1}  \E \Bigg[ \sum_{t=1}^T I_{t,J}I_{t,J^\prime} W_{t,J, J^\prime}(A_{t,J},&A_{t,J^\prime},H_t; \eta^\star,\rho^\star) Y_{t,1,J} \nonumber \\
&\begin{pmatrix}
  g_t(H_t) \\
  (1-A_{t,J})(A_{t,J^\prime} - \tilde {p}^\star_t (1 \given S_t) ) f_t (S_t)
\end{pmatrix} \Bigg]
\end{align*}

Then standard statistical arguments can be used to show that $\sqrt{n}(\hat\alpha-\alpha^\prime, \hat\beta-\beta^\prime)$ converges in distribution to a normal, mean zero, random vector with variance-covariance matrix given by:
\begin{equation*}
    \left\{ \E \left[\dot U_W(\eta^\star, \rho^\star) \right]\right\}^{-1} \Sigma_W(\alpha^\prime,\beta^\prime;\eta^\star, \rho^\star) \left\{ \E \left[\dot U_W(\eta^\star, \rho^\star) \right]\right\}^{-1}
\end{equation*}
where
\begin{align*}
    \Sigma_W(\alpha,\beta;\eta, \rho) = \E \Big[\Big(U_W(\alpha,\beta;\eta, \rho) + &\Sigma_{W,D}(\alpha,\beta;\eta, \rho) \left\{\E[\dot U_D(\eta)] \right\}^{-1} U_D(\eta) + \nonumber \\ &\Sigma_{W,N}(\alpha,\beta;\eta, \rho)\left\{\E[\dot U_N(\rho)] \right\}^{-1} U_N(\rho) \Big)^ {\otimes 2} \Big]
\end{align*}
with
\begin{align*}
    \Sigma_{W,D}&(\alpha,\beta ; \eta, \rho)
    = \E \Big[ \sum_{t=1}^T \left( Y_{t,1,J} - g_t(H_t)^\top \alpha -  (1-A_{t,J})(A_{t,J^\prime} - \tilde p_t^\star (1 \given S_t;\rho) ) f_t (S_t)^\top \beta \right)I_{t,J}I_{t,J^\prime} \nonumber \\ &W_{t,J, J^\prime}(A_{t,J},A_{t,J^\prime},H_t; \eta,\rho)
    \begin{pmatrix}
  g_t(H_t) \\
  (1-A_{t,J})(A_{t,J^\prime} - \tilde {p}^\star_t (1 \given S_t;\rho) ) f_t (S_t)
\end{pmatrix}\left(\frac{d \log p^\star_t(A_{t,J^\prime}|H_t;\eta)}{d \eta} \right)^\top \Big],
\end{align*}
and
\begin{align*}
    \Sigma_{W,N}&(\alpha,\beta;\eta, \rho)
    = \E \Big[ \sum_{t=1}^T \left( Y_{t,1,J} - g_t(H_t)^\top \alpha -  (1-A_{t,J})(A_{t,J^\prime} - \tilde p_t^\star (1 \given S_t;\rho) ) f_t (S_t)^\top \beta \right)I_{t,J}I_{t,J^\prime} \nonumber \\ &W_{t,J, J^\prime}(A_{t,J},A_{t,J^\prime},H_t; \eta,\rho)
    \begin{pmatrix}
  g_t(H_t) \\
  (1-A_{t,J})(A_{t,J^\prime} - \tilde {p}^\star_t (1 \given S_t;\rho) ) f_t (S_t)
\end{pmatrix}\left(\frac{d \log \tilde p^\star_t(A_{t,J^\prime}|S_t;\rho)}{d \rho} \right)^\top \Big] \nonumber \\
&+\E \Big[ \sum_{t=1}^T \left( Y_{t,1,J} - g_t(H_t)^\top \alpha -  (1-A_{t,J})(A_{t,J^\prime} - \tilde p_t^\star (1 \given S_t;\rho) ) f_t (S_t)^\top \beta \right)I_{t,J}I_{t,J^\prime} \nonumber \\ &W_{t,J, J^\prime}(A_{t,J},A_{t,J^\prime},H_t; \eta,\rho)
    \begin{pmatrix}
  \textbf{0} \\
  -(1-A_{t,J})\tilde {p}^\star_t (1 \given S_t;\rho)  f_t (S_t)
\end{pmatrix}\left(\frac{d \log \tilde p^\star_t(1|S_t;\rho)}{d \rho} \right)^\top \Big] \nonumber \\
&+\E \Big[ \sum_{t=1}^T (1-A_{t,J}) \tilde p_t^\star (1 \given S_t;\rho)  f_t (S_t)^\top \beta I_{t,J}I_{t,J^\prime}W_{t,J, J^\prime}(A_{t,J},A_{t,J^\prime},H_t; \eta,\rho)  \nonumber \\ &\begin{pmatrix}
  g_t(H_t) \\
  (1-A_{t,J})(A_{t,J^\prime} - \tilde {p}^\star_t (1 \given S_t;\rho) ) f_t (S_t)
\end{pmatrix}\left(\frac{d \log \tilde p^\star_t(1|S_t;\rho)}{d \rho} \right)^\top \Big]
\end{align*}

In our simulation, an individual's randomization probabilities only depends on their observed history, then $\tilde p_t^\star (1 \mid S_{t,j^\prime}) = \tilde p_t (1 \mid S_{t,j^\prime})$. Since the data is from an MRT (we know $p_t$) and we pre-specify (not estimate) $\tilde p_t$, then $\Sigma_W = \E \left[ \left(U_W(\alpha,\beta)  \right)^ {\otimes 2} \right] $ greatly simplifying the variance-covaraince matrix.

A consistent estimator of the variance-covariance matrix is given by:
\begin{equation}
    \left\{ \P_n \left[\dot U_W(\hat\eta, \hat\rho) \right]\right\}^{-1} \hat\Sigma_W(\hat\alpha,\hat\beta;\hat\eta, \hat\rho) \left\{ \P_n \left[\dot U_W(\hat\eta, \hat\rho) \right]\right\}^{-1}
\end{equation}
where
\begin{align*}
    \hat\Sigma_W(\alpha,\beta;\eta, \rho) = \P_n \Big[\Big(U_W(\alpha,\beta;\eta, \rho) + &\hat\Sigma_{W,D}(\alpha,\beta;\eta, \rho) \left\{\P_n[\dot U_D(\eta)] \right\}^{-1} U_D(\eta) + \nonumber \\ &\hat\Sigma_{W,N}(\alpha,\beta;\eta, \rho)\left\{\P_n[\dot U_N(\rho)] \right\}^{-1} U_N(\rho) \Big)^ {\otimes 2} \Big]
\end{align*}
with
\begin{align*}
    \hat\Sigma_{W,D}(\alpha,\beta &; \eta, \rho)
    = \P_n \Big[ \sum_{t=1}^T \left( Y_{t,1,J} - g_t(H_t)^\top \alpha -  (1-A_{t,J})(A_{t,J^\prime} - \tilde p_t^\star (1 \given S_t;\rho) ) f_t (S_t)^\top \beta \right)I_{t,J}I_{t,J^\prime} \nonumber \\ &W_{t,J, J^\prime}(A_t,H_t; \eta,\rho)
    \begin{pmatrix}
  g_t(H_t) \\
  (1-A_{t,J})(A_{t,J^\prime} - \tilde {p}^\star_t (1 \given S_t;\rho) ) f_t (S_t)
\end{pmatrix}\left(\frac{d \log p^\star_t(A_{t,J^\prime}|H_t;\eta)}{d \eta} \right)^\prime \Big],
\end{align*}
and
\begin{align*}
    \hat\Sigma_{W,N}&(\alpha,\beta;\eta, \rho)
    = \P_n \Big[ \sum_{t=1}^T \left( Y_{t,1,J} - g_t(H_t)^\top \alpha -  (1-A_{t,J})(A_{t,J^\prime} - \tilde p_t^\star (1 \given S_t;\rho) ) f_t (S_t)^\top \beta \right)I_{t,J}I_{t,J^\prime} \nonumber \\ &W_{t,J, J^\prime}(A_{t,J},A_{t,J^\prime},H_t; \eta,\rho)
    \begin{pmatrix}
  g_t(H_t) \\
  (1-A_{t,J})(A_{t,J^\prime} - \tilde {p}^\star_t (1 \given S_t;\rho) ) f_t (S_t)
\end{pmatrix}\left(\frac{d \log \tilde p^\star_t(A_{t,J^\prime}|S_t;\rho)}{d \rho} \right)^\top \Big] \nonumber \\
&+\P_n \Big[ \sum_{t=1}^T \left( Y_{t,1,J} - g_t(H_t)^\top \alpha -  (1-A_{t,J})(A_{t,J^\prime} - \tilde p_t^\star (1 \given S_t;\rho) ) f_t (S_t)^\top \beta \right)I_{t,J}I_{t,J^\prime} \nonumber \\ &W_{t,J, J^\prime}(A_{t,J},A_{t,J^\prime},H_t; \eta,\rho)
    \begin{pmatrix}
  \textbf{0} \\
  -(1-A_{t,J})\tilde {p}^\star_t (1 \given S_t;\rho)  f_t (S_t)
\end{pmatrix}\left(\frac{d \log \tilde p^\star_t(1|S_t;\rho)}{d \rho} \right)^\top \Big] \nonumber \\
&+\P_n \Big[ \sum_{t=1}^T (1-A_{t,J}) \tilde p_t^\star (1 \given S_t;\rho)  f_t (S_t)^\top \beta I_{t,J}I_{t,J^\prime}W_{t,J, J^\prime}(A_{t,J},A_{t,J^\prime},H_t; \eta,\rho)  \nonumber \\ &\begin{pmatrix}
  g_t(H_t) \\
  (1-A_{t,J})(A_{t,J^\prime} - \tilde {p}^\star_t (1 \given S_t;\rho) ) f_t (S_t)
\end{pmatrix}\left(\frac{d \log \tilde p^\star_t(1|S_t;\rho)}{d \rho} \right)^\top \Big]
\end{align*}

It remains to show that $\beta^\prime = \beta^{\star\star}$. Since $\E [U_W(\alpha^\prime,\beta^\prime;\eta^\star,\rho^\star)]=0$,
\begin{align*}
    0 &= \E \sum_{t=1}^T \left( Y_{t,1,J} - g_t(H_t)^\top\alpha^\prime -  (1-A_{t,J}) (A_{t,J^\prime} - \tilde p_t^\star (1 \given S_t;\rho^\star) ) f_t (S_t)^\top \beta^\prime \right) I_{t,J}I_{t,J^\prime} \nonumber \\
    &  ~~~~~~~~~~~~~~~~~~~~  W_{t,J, J^\prime}(A_{t,J},A_{t,J^\prime},H_t; \eta^\star,\rho^\star)(1-A_{t,J})(A_{t,J^\prime} - \tilde {p}^\star_t (1 \given S_t;\rho^\star) ) f_t (S_t) \nonumber \\
    &=  \E \sum_{t=1}^T \left(\E \left[Y_{t,1} \given{A_{t,J},A_{t,J^\prime}, H_t ,I_{t,J}I_{t,J^\prime}=1} \right]- g_t(H_t)^\top\alpha^\prime-(1-A_{t,J}) (A_{t,J^\prime} - \tilde p_t^\star (1 \given S_t;\rho^\star) ) f_t (S_t)^\top \beta^\prime \right) \nonumber \\
    &  ~~~~~~~~~~~~~~~~~~~~  I_{t,J}I_{t,J^\prime}W_{t,J, J^\prime}(A_{t,J},A_{t,J^\prime},H_t; \eta^\star,\rho^\star)(1-A_{t,J})(A_{t,J^\prime} - \tilde {p}^\star_t (1 \given S_t;\rho^\star) ) f_t (S_t) \nonumber \\
    &= \E \sum_{t=1}^T \sum_{a^\prime \in \{0,1\}} \Big(\E \left[Y_{t,1} \given{A_{t,J}=0,A_{t,J^\prime}=a^\prime, H_t ,I_{t,J}I_{t,J^\prime}=1} \right]- g_t(H_t)^\top\alpha^\prime- \nonumber \\
    &  ~~~~~~~~~~~~~~~~~~~~  (a^\prime - \tilde p_t^\star (1 \given S_t;\rho^\star) ) f_t (S_t)^\top \beta^\prime \Big) I_{t,J}I_{t,J^\prime}\tilde p_t (0,a^\prime \given S_t;\rho^\star)(a^\prime - \tilde {p}^\star_t (1 \given S_t;\rho^\star) ) f_t (S_t)
\end{align*}

where the last equality averages out over $A_{t,J^\prime}$. The above simplifies to:

\begin{align*}
    0 &= \E \sum_{t=1}^T \sum_{a^\prime \in \{0,1\}} \Big(\E \left[Y_{t,1} \given{A_{t,J}=0,A_{t,J^\prime}=a^\prime, H_t ,I_{t,J}I_{t,J^\prime}=1} \right]- g_t(H_t)^\top\alpha^\prime- \nonumber \\
    &  ~~~~~~~~~~~~~~~~~~~~  (a^\prime - \tilde p_t^\star (1 \given S_t;\rho^\star) ) f_t (S_t)^\top \beta^\prime \Big) I_{t,J}I_{t,J^\prime}\tilde p_t (0,a^\prime \given S_t;\rho^\star)(a^\prime - \tilde {p}^\star_t (1 \given S_t;\rho^\star) ) f_t (S_t) \nonumber \\
    &= \E \sum_{t=1}^T  \Big(\E \left[Y_{t,1} \given{A_{t,J}=0,A_{t,J^\prime}=1, H_t ,I_{t,J}I_{t,J^\prime}=1} \right]- g_t(H_t)^\top\alpha^\prime- \nonumber \\
    &  ~~~~~~~~~~~~~~~~~~~~  (1 - \tilde p_t^\star (1 \given S_t;\rho^\star) ) f_t (S_t)^\top \beta^\prime \Big) I_{t,J}I_{t,J^\prime}\tilde p_t (0,1 \given S_t;\rho^\star)(1 - \tilde {p}^\star_t (1 \given S_t;\rho^\star) ) f_t (S_t) \nonumber \\
    & ~~~~ +\E \sum_{t=1}^T \Big(\E \left[Y_{t,1} \given{A_{t,J}=0,A_{t,J^\prime}=0, H_t ,I_{t,J}I_{t,J^\prime}=1} \right]- g_t(H_t)^\top\alpha^\prime- \nonumber \\
    &  ~~~~~~~~~~~~~~~~~~~~  ( - \tilde p_t^\star (1 \given S_t;\rho^\star) ) f_t (S_t)^\top \beta^\prime \Big) I_{t,J}I_{t,J^\prime}\tilde p_t (0,0 \given S_t;\rho^\star)( - \tilde {p}^\star_t (1 \given S_t;\rho^\star) ) f_t (S_t) \nonumber \\
    &= \E \sum_{t=1}^T \Big(\E \left[Y_{t,1} \given{A_{t,J}=0,A_{t,J^\prime}=1, H_t ,I_{t,J}I_{t,J^\prime}=1} \right] - \E \left[Y_{t,1} \given{A_{t,J}=0,A_{t,J^\prime}=0, H_t ,I_{t,J}I_{t,J^\prime}=1} \right]\nonumber \\
    &  ~~~~~~~~~~~~~~~~~~~~  -f_t (S_t)^\top \beta^\prime \Big)  f_t (S_t) \gamma(\eta^\star,\rho^\star) I_{t,J}I_{t,J^\prime}
\end{align*}

where $\gamma(\eta^\star,\rho^\star) = \tilde{p}_t (0,1 \mid S_t)(1- \tilde {p}^\star_t (1 \given S_t;\rho^\star)) = \tilde{p}_t (0,0 \mid S_t) \tilde {p}^\star_t (1 \given S_t;\rho^\star) $. From this we obtain:
\begin{align*}
   0 = \E \sum_{t=1}^T  f_t (S_t) &\gamma(\eta^\star,\rho^\star) I_{t,J}I_{t,J^\prime}   \Big(\E \Big[\E \left[Y_{t,1} \given{A_{t,J}=0,A_{t,J^\prime}=1, H_t ,I_{t,J}I_{t,J^\prime}=1} \right] - \nonumber \\
   &\E \left[Y_{t,1} \given{A_{t,J}=0,A_{t,J^\prime}=0, H_t ,I_{t,J}I_{t,J^\prime}=1} \right] \mid S_t,I_{t,J}I_{t,J^\prime}=1\Big]  -f_t (S_t)^\top \beta^\prime \Big)
\end{align*}

Thus
\begin{align}
\label{eq:estimand_beta}
    \beta^\prime =\left[\E \dot U_W(\eta^\star,\rho^\star) \right]_{(2,2)}^{-1} \E \Bigg[\sum_{t=1}^T  f_t (S_t) &\gamma(\eta^\star,\rho^\star) I_{t,J}I_{t,J^\prime}  \E \Big[\E \left[Y_{t,1} \given{A_{t,J}=0,A_{t,J^\prime}=1, H_t ,I_{t,J}I_{t,J^\prime}=1} \right] - \nonumber \\
   &\E \left[Y_{t,1} \given{A_{t,J}=0,A_{t,J^\prime}=0, H_t ,I_{t,J}I_{t,J^\prime}=1} \right] \mid S_t,I_{t,J}I_{t,J^\prime}=1\Big]  \Bigg]
\end{align}

where
\[
\left[\E \dot U_W(\eta^\star,\rho^\star) \right]_{(2,2)} =\E \sum_{t=1}^T  f_t (S_t)f_t (S_t)^\top \gamma(\eta^\star,\rho^\star) I_{t,J}I_{t,J^\prime}
\]
\end{proof}

\section{Connection to a Semiparametric Efficient Estimator}
\label{sec:semipar}

A special case of both effects is when $S_t$ is set to the observed history~$H_t$ and~$\Delta = 1$.  In this case, estimators for this fully conditional case can be derived using techniques from~\cite{Robins1994} based on semiparametric efficiency theory~\citep{Newey1990,Tsiatis2007} under a lack of interference assumption.  Proofs can be found in Appendix~\ref{app:semipareff}.

\begin{lemma}
Under the semiparametric model~\eqref{ass:directeffect}, Assumption~\ref{consistency}, and the stronger \emph{lack of interference} assumption, the semiparametric efficient score~$S_{\text{eff}} (\beta)$ for $\beta$ is
\begin{equation}
\label{eq:effscore}
\frac{1}{G} \sum_{j=1}^G \sum_{t=1}^T (Y_{t,1,j} - \mu (H_{t,j}) - (A_{t,j} - p_t (1 \mid H_{t,j})) f_t (H_{t,j})^\top \beta ) K_{t,j} (A_{t,j} - p_t(1 \mid H_{t,j})) f(H_{t,j}),
\end{equation}
where
\begin{align*}
\mu_t (H_{t,j}) = \mathbb{E} &\left[ Y_{t,1,j} \mid H_{t,j} \right], \quad \text{and} \quad \sigma^2_{t+1} (H_{t,j}, A_{t,j}) = \text{Var} \left( Y_{t,1,j} \mid H_{t,j}, A_{t,j} \right) \\
K_{t,j} = &\bigg[ \frac{1}{\sigma^2_{t+1,j} (H_{t,j}, 1)} + p_t ( 1 \mid H_{t,j}) \left[  \frac{p_t (1 \mid H_{t,j} )}{\sigma^2_{t+1,j} (H_{t,j}, 1)} + \frac{1-p_t (1 \mid H_{t,j} )}{\sigma^2_{t+1,j} (H_{t,j}, 0)} \right] \times \\
&\left( \frac{1}{\sigma^2_{t+1,j} (H_{t,j}, 1)}  - \frac{1}{\sigma^2_{t+1,j} (H_{t,j}, 0)} \right)
 \bigg].
\end{align*}
\end{lemma}

Semiparametric efficiency theory states that the solution $\hat \beta$ to $\mathbb{P}_n \left[ S_{\text{eff}} (\beta) \right] = 0$ achieves the semiparametric efficiency bound~\citep{Newey1990}.
% \cite{Qian2021} proved a similar result for the causal excursion effect on the relative risk scale for binary outcomes.  Unlike~\cite{Qian2021}, the efficient score here requires estimates of the conditional variance, which may be difficult to obtain.
Corollary~\ref{cor:semiparametricconnection} motivates the C-WCLS criterion for the direct effect given by~\eqref{eq:directwcls} from the semiparametric efficiency perspective under particular working homoskedastic assumptions on the conditional variance~$\sigma^2_{t+1} (H_{t,j}, A_{t,j})$ and a working model~$g_t (H_{t,j})^\top \alpha$ for the unknown quantity~$\mu(H_{t,j})$.
% Making the homoskedastic assumption helps avoid the conditional variance estimation problem.
While this establishes a connection between the C-WCLS criterion and semiparametric efficiency scores, more work on efficiency theory for causal excursions effects is considered important future work.

\begin{corollary}
\label{cor:semiparametricconnection}
Assuming $\text{Var}(Y_{t,1,j} \mid H_{t,j}, A_{t,j}) := \sigma^2_{t+1}(H_{t,j})$, i.e., is constant in the second argument~$A_{t,j}$, the weight~$K_{t,j}$ is equal to $(\sigma^2_{t+1}(H_{t,j}))^{-1}$ and the semiparametric efficient score~\eqref{eq:effscore} weights each decision time by the corresponding conditional variance.  Under the further assumption that this variance does not depend on the history, i.e., $\sigma^2_{t+1} (H_{t,j}) = \sigma^2$, and a working model for the conditional mean $\mu (H_{t,j}) := g (H_{t,j})^\top \alpha$, criterion~\eqref{eq:directwcls} is equivalent to the semiparametric efficient score for the fully conditional effect.
\end{corollary}

\subsection{Proof of semiparametric efficiency}
\label{app:semipareff}

In this section, we assume lack of interference and therefore the potential outcomes can be written to only depend on one's observed history.  Then we consider a semiparametric model characterized by the following assumptions:
\begin{assumption}
For all $1 \leq t \leq T$, $E[ Y_{t,1,J} (\bar A_{t-1,J}, 0) \mid H_{t,J}, A_{t,J} ] = E[ Y_{t,1,J} (\bar A_{t-1,J}, 0) \mid H_{t,J} ]$
\end{assumption}
\begin{assumption}
Assume that there exists a function $\gamma()$ and a true parameter $\psi_0 \in \mathbb{R}^p$, such that for any $1 \leq t \leq T$,
$$
\mathbb{E} \left[ Y_{t,1,J} (\bar A_{t-1,J}, a_t) \mid \bar z_t, \bar a_t \right] - \mathbb{E} \left[ Y_{t,1,J} (\bar A_{t-1,J}, 0) \mid \bar z_t, \bar a_t \right] = \gamma(t+1, \bar z_t, \bar a_t; \psi)
$$
\end{assumption}
We next gather the definitions necessary for defining the semiparametric efficient score:
\begin{itemize}
\item The longitudinal data is $O_1, A_1, Y_{2}, O_2, A_2, \ldots, O_T, A_T, Y_{t,1}$ where $O_t$ is the time-varying covariates on all individuals in the cluster, $A_t$ is the treatment assignments for the cluster, and $Y_{t,1}$ is the set of proximal outcomes on the cluster
\item $Z_{t,j} = (Y_{t,j}, O_{t,j})$
\item $H_{t,j} = (\bar A_{t-1,j}, \bar Z_{t,j})$
\item $V_{t,j} = (H_{t,j}, A_{t,j})$
\item $U_{t+1,j} (\psi) = Y_{t,1,J} - \gamma(t+1, \bar z_t, \bar a_t; \psi)$
\item $\dot{U}_{t+1,j} (\psi) = U_{t+1,j} - \mathbb{E} \left[ U_{t+1, j} \mid H_{t,j} \right]$
\item $W_{t,j} = \text{Var} \left( U_{t+1,j} (\psi_0) \mid V_{t,j} \right)^{-1}$
\end{itemize}

Then by~\cite[Lemma I.8]{Qian2021}, a general form of the efficient score is
$$
S_{\text{eff}} (\psi_0) = - \frac{1}{G} \sum_{j=1}^G \sum_{t=1}^T \rho_{t,j} \dot{U}_{t+1,j} (\psi_0)
$$
where
$$
\rho_{t,j} = \left[ \mathbb{E} \left[ \frac{\partial U_{t+1,j}}{\partial \psi} \mid V_{t,j} \right] - \mathbb{E} \left[ \frac{\partial U_{t+1,j}}{\partial \psi} \mid H_{t,j} \right] \mathbb{E} \left( W_{t,j} \mid H_{t,j} \right)^{-1} \right] W_{t,j}
$$
Note that $\mathbb{E} \left[ \rho_{t,j} \mid H_t \right] = 0$.  Therefore by~\cite[Lemma I.1]{Qian2021} we have
$$
\rho_{t,j} = \left( \rho(A_{t,j} = 1) - \rho(A_{t,j} = 1) \right) (A_{t,j} - p_t (1 \mid H_{t,j}))
$$
where  $\rho(A_{t,j} = a)$ denotes $\rho_{t,j}$ evaluated at $A_{t,j} = a$.

We now calculate these terms based on the above notation. Under $\gamma(t+1, \bar z_t, \bar a_t; \psi_0) = A_{t,j} f(H_{t,j})^\top \psi$, we have
$$
\frac{\partial U_{t+1,j} (\psi_0)}{\partial \psi} = - A_{t,j} f(H_t), \quad \text{and} \quad
\dot{U}_{t+1,j} (\psi) = Y_{t,1,J} - \mu_t (H_t) + (A_{t,j}-p_t(1 \mid H_t)) f_t(H_{t,j})^\top \psi
$$
and hence we have
\begin{align*}
\mathbb{E} \left[ \frac{\partial U_{t+1,j} (\psi_0)}{\partial \psi} \mid H_{t,j}, A_{t,j} = 1 \right] &= - f(H_{t,j}) \\
\mathbb{E} \left[ \frac{\partial U_{t+1,j} (\psi_0)}{\partial \psi} \mid H_{t,j}, A_{t,j} = 0 \right] &= 0 \\
\text{Var} \left( U_{t+1,j} (\psi_0) \mid V_{t,j} \right) &=
\text{Var} \left( Y_{t,1,J} \mid V_{t,j} \right) =: \sigma^2_{t+1,j} (H_{t,j}, A_{t,j}) \\
\end{align*}
Then
$$
\mathbb{E} \left[ W_{t,j} \mid H_{t,j} \right] = \frac{p_t (1 \mid H_{t,j} )}{\sigma^2_{t+1,j} (H_{t,j}, 1)} + \frac{1-p_t (1 \mid H_{t,j} )}{\sigma^2_{t+1,j} (H_{t,j}, 0)}
$$
and we can express
\begin{align*}
\rho(A_{t,j} = 1) &= - \left( 1 -p_t ( 1 \mid H_{t,j}) \left[  \frac{p_t (1 \mid H_{t,j} )}{\sigma^2_{t+1,j} (H_{t,j}, 1)} + \frac{1-p_t (1 \mid H_{t,j} )}{\sigma^2_{t+1,j} (H_{t,j}, 0)} \right] \right) \frac{f(H_{t,j})}{\sigma^2_{t+1,j} (H_{t,j}, 1)} \\
\rho(A_{t,j} = 0) &= - \left( 0 -p_t ( 1 \mid H_{t,j}) \left[  \frac{p_t (1 \mid H_{t,j} )}{\sigma^2_{t+1,j} (H_{t,j}, 1)} + \frac{1-p_t (1 \mid H_{t,j} )}{\sigma^2_{t+1,j} (H_{t,j}, 0)} \right] \right) \frac{f(H_{t,j})}{\sigma^2_{t+1,j} (H_{t,j}, 1)}
\end{align*}
Therefore $\rho_{t,j}$ is given by
\begin{align*}
\bigg[ \frac{1}{\sigma^2_{t+1,j} (H_{t,j}, 1)} + p_t ( 1 \mid H_{t,j}) \left[  \frac{p_t (1 \mid H_{t,j} )}{\sigma^2_{t+1,j} (H_{t,j}, 1)} + \frac{1-p_t (1 \mid H_t )}{\sigma^2_{t+1,j} (H_{t,j}, 0)} \right] \times \\
\left( \frac{1}{\sigma^2_{t+1,j} (H_{t,j}, 1)}  - \frac{1}{\sigma^2_{t+1,j} (H_{t,j}, 0)} \right)
 \bigg] \times \left( A_{t,j} - p_t (1 \mid H_{t,j}) \right) f(H_{t,j}).
\end{align*}
Moreover, under the simplifying assumption $\sigma^2_{t+1,j} (H_{t,j}, a) = \sigma^2_{t+1,j} (H_{t,j})$ we have
$$
\rho_{t,j} = \frac{1}{\sigma^2_{t+1,j} (H_t)} \times \left( A_{t,j} - p_t (1 \mid H_{t,j}) \right) f(H_{t,j}).
$$
Under the even stronger assumption $\sigma^2_{t+1,j} (H_{t,j}) := \sigma^2$ we have
\begin{align*}
S_{\text{eff}} (\psi_0) =&  \frac{\sigma^2}{G} \sum_{j=1}^G \sum_{t=1}^T \left(Y_{t,1,J} - \mu_t (H_{t,j}) - (A_{t,j} - p_t (1 \mid H_{t,j})) f_t (H_{t,j})^\top \beta \right) \times \\
&\times (A_{t,j} - p_t (1 \mid H_{t,j})) f_t (H_{t,j}).
\end{align*}

\section{Code to Replicate Simulation and Case Study Results}
The R code used to generate the simulation experiments and case study results in this paper can be obtained at \verb"https://github.com/Herashi/MRT-mHealthModeration".
\end{document}